\pdfoutput=1
\documentclass[epj]{svjour}
\usepackage{ifthen} 
\usepackage{multirow}
\usepackage{pdflscape}
\usepackage{float}
\usepackage{booktabs}
\usepackage{rotating}
\usepackage{amsmath}
\usepackage{amssymb}
\usepackage{rotating}
\usepackage{verbatim}  
\usepackage[flushleft]{threeparttable}
\usepackage[usenames,dvipsnames]{xcolor}
\definecolor{taplum}{rgb}{0.67843, 0.49804, 0.65882}
\usepackage{bm} 
\usepackage{float} 
\usepackage{subfig} 
\newboolean{pdflatex}
\setboolean{pdflatex}{true} 
\usepackage{epstopdf}
\newboolean{articletitles}
\setboolean{articletitles}{true} 
\newboolean{uprightparticles}
\setboolean{uprightparticles}{false} 
\newboolean{inbibliography}
\setboolean{inbibliography}{false} 

\usepackage[top=1in, bottom=1.25in, left=1in, right=1in]{geometry}

\usepackage{microtype}
\usepackage{lineno}  
\usepackage{xspace} 
\usepackage{caption} 

\usepackage{graphicx}  
\usepackage{color}
\usepackage{colortbl}
\graphicspath{{./figs/}} 

\usepackage{amsmath} 
\usepackage{amssymb}
\usepackage{amsfonts}
\usepackage{upgreek} 

\newcommand*\patchAmsMathEnvironmentForLineno[1]{%
\expandafter\let\csname old#1\expandafter\endcsname\csname #1\endcsname
\expandafter\let\csname oldend#1\expandafter\endcsname\csname
end#1\endcsname
 \renewenvironment{#1}%
   {\linenomath\csname old#1\endcsname}%
   {\csname oldend#1\endcsname\endlinenomath}%
}
\newcommand*\patchBothAmsMathEnvironmentsForLineno[1]{%
  \patchAmsMathEnvironmentForLineno{#1}%
  \patchAmsMathEnvironmentForLineno{#1*}%
}
\AtBeginDocument{%
\patchBothAmsMathEnvironmentsForLineno{equation}%
\patchBothAmsMathEnvironmentsForLineno{align}%
\patchBothAmsMathEnvironmentsForLineno{flalign}%
\patchBothAmsMathEnvironmentsForLineno{alignat}%
\patchBothAmsMathEnvironmentsForLineno{gather}%
\patchBothAmsMathEnvironmentsForLineno{multline}%
\patchBothAmsMathEnvironmentsForLineno{eqnarray}%
}

\usepackage{hyperref}    
\usepackage[all]{hypcap} 

\usepackage{xspace} 
\usepackage{upgreek}

\def\lhcb {\mbox{LHCb}\xspace}

\def\lhc    {\mbox{LHC}\xspace}

\def\MagUp {\mbox{\em Mag\kern -0.05em Up}\xspace}

\ifthenelse{\boolean{uprightparticles}}%
{

 \def\Ppi         {\ensuremath{\uppi}\xspace}

 \def\Ppsi        {\ensuremath{\uppsi}\xspace}

 \def\PDelta      {\ensuremath{\Delta}\xspace}                 
 \def\PXi      {\ensuremath{\Xi}\xspace}                 
 \def\PLambda      {\ensuremath{\Lambda}\xspace}                 
 \def\PSigma      {\ensuremath{\Sigma}\xspace}                 
 \def\POmega      {\ensuremath{\Omega}\xspace}                 
 \def\PUpsilon      {\ensuremath{\Upsilon}\xspace}                 
 
 \def\PB      {\ensuremath{\mathrm{B}}\xspace}                 
                  
 \def\PD      {\ensuremath{\mathrm{D}}\xspace}

 \def\PJ      {\ensuremath{\mathrm{J}}\xspace}                 
 \def\PK      {\ensuremath{\mathrm{K}}\xspace}

 \def\PV      {\ensuremath{\mathrm{V}}\xspace}

 \def\Pb      {\ensuremath{\mathrm{b}}\xspace}                 
 \def\Pc      {\ensuremath{\mathrm{c}}\xspace}

 \def\Pi      {\ensuremath{\mathrm{i}}\xspace}

 \def\Pp      {\ensuremath{\mathrm{p}}\xspace}                 
 \def\Pq      {\ensuremath{\mathrm{q}}\xspace}                 
                  
 \def\Ps      {\ensuremath{\mathrm{s}}\xspace}

}
{

 \def\Ppi         {\ensuremath{\pi}\xspace}

 \def\Ppsi        {\ensuremath{\psi}\xspace}                 
                  
 \mathchardef\PDelta="7101
 \mathchardef\PXi="7104
 \mathchardef\PLambda="7103
 \mathchardef\PSigma="7106
 \mathchardef\POmega="710A
 \mathchardef\PUpsilon="7107
                  
 \def\PB      {\ensuremath{B}\xspace}                 
                  
 \def\PD      {\ensuremath{D}\xspace}

 \def\PJ      {\ensuremath{J}\xspace}                 
 \def\PK      {\ensuremath{K}\xspace}

 \def\PV      {\ensuremath{V}\xspace}

 \def\Pb      {\ensuremath{b}\xspace}                 
 \def\Pc      {\ensuremath{c}\xspace}

 \def\Pi      {\ensuremath{i}\xspace}

 \def\Pp      {\ensuremath{p}\xspace}                 
 \def\Pq      {\ensuremath{q}\xspace}                 
                  
 \def\Ps      {\ensuremath{s}\xspace}

}

\makeatletter
\ifcase \@ptsize \relax
  \newcommand{\miniscule}{\@setfontsize\miniscule{4}{5}}
\or
  \newcommand{\miniscule}{\@setfontsize\miniscule{5}{6}}
\or
  \newcommand{\miniscule}{\@setfontsize\miniscule{5}{6}}
\fi
\makeatother

\DeclareRobustCommand{\optbar}[1]{\shortstack{{\miniscule (\rule[.5ex]{1.25em}{.18mm})}
  \\ [-.7ex] $#1$}}

\def\W      {{\ensuremath{\PW}}\xspace}

\def\quark     {{\ensuremath{\Pq}}\xspace}
\def\quarkbar  {{\ensuremath{\overline \quark}}\xspace}

\def\squark    {{\ensuremath{\Ps}}\xspace}

\def\cquark    {{\ensuremath{\Pc}}\xspace}

\def\bquark    {{\ensuremath{\Pb}}\xspace}
\def\bquarkbar {{\ensuremath{\overline \bquark}}\xspace}
\def\bbbar     {{\ensuremath{\bquark\bquarkbar}}\xspace}

\def\pion   {{\ensuremath{\Ppi}}\xspace}

\def\pip    {{\ensuremath{\pion^+}}\xspace}
\def\pim    {{\ensuremath{\pion^-}}\xspace}

\def\kaon    {{\ensuremath{\PK}}\xspace}
  \def\Kbar    {{\kern 0.2em\overline{\kern -0.2em \PK}{}}\xspace}

\def\KorKbar    {\kern 0.18em\optbar{\kern -0.18em K}{}\xspace}

\def\Kp      {{\ensuremath{\kaon^+}}\xspace}
\def\Km      {{\ensuremath{\kaon^-}}\xspace}

\def\Kstarz  {{\ensuremath{\kaon^{*0}}}\xspace}
\def\Kstarzb {{\ensuremath{\Kbar{}^{*0}}}\xspace}

  \def\Dbar    {{\kern 0.2em\overline{\kern -0.2em \PD}{}}\xspace}
\def\D       {{\ensuremath{\PD}}\xspace}

\def\DorDbar    {\kern 0.18em\optbar{\kern -0.18em D}{}\xspace}

\def\Dp      {{\ensuremath{\D^+}}\xspace}

\def\Dsp     {{\ensuremath{\D^+_\squark}}\xspace}

\def\Bbar    {{\ensuremath{\kern 0.18em\overline{\kern -0.18em \PB}{}}}\xspace}

\def\BorBbar    {\kern 0.18em\optbar{\kern -0.18em B}{}\xspace}

\def\jpsi     {{\ensuremath{{\PJ\mskip -3mu/\mskip -2mu\Ppsi\mskip 2mu}}}\xspace}

  \def\Y#1S{\ensuremath{\PUpsilon{(#1S)}}\xspace}

\def\proton      {{\ensuremath{\Pp}}\xspace}

\def\Deltares    {{\ensuremath{\PDelta}}\xspace}

\def\Lz          {{\ensuremath{\PLambda}}\xspace}
\def\Lbar        {{\ensuremath{\kern 0.1em\overline{\kern -0.1em\PLambda}}}\xspace}
\def\LorLbar    {\kern 0.18em\optbar{\kern -0.18em \PLambda}{}\xspace}

\def\Lb      {{\ensuremath{\Lz^0_\bquark}}\xspace}

\def\Lc      {{\ensuremath{\Lz^+_\cquark}}\xspace}
\def\Lcbar   {{\ensuremath{\Lbar{}^-_\cquark}}\xspace}

\def\Xibm    {{\ensuremath{\Xires^-_\bquark}}\xspace}

\def\Xicz    {{\ensuremath{\Xires^0_\cquark}}\xspace}
\def\Xicp    {{\ensuremath{\Xires^+_\cquark}}\xspace}

\def\Xicbarz {{\ensuremath{\Xiresbar{}_\cquark^0}}\xspace}

\def\to                 {\ensuremath{\rightarrow}\xspace}

\def\eps   {{\ensuremath{\varepsilon}}\xspace}

\def\CP                {{\ensuremath{C\!P}}\xspace}
\def\CPT               {{\ensuremath{C\!PT}}\xspace}

\def\AT#1     {\ensuremath{A_{\mathrm{T}}^{#1}}\xspace}           

\def\C#1      {\ensuremath{\mathcal{C}_{#1}}\xspace}                       
\def\Cp#1     {\ensuremath{\mathcal{C}_{#1}^{'}}\xspace}                    
\def\Ceff#1   {\ensuremath{\mathcal{C}_{#1}^{\mathrm{(eff)}}}\xspace}        
\def\Cpeff#1  {\ensuremath{\mathcal{C}_{#1}^{'\mathrm{(eff)}}}\xspace}       
\def\Ope#1    {\ensuremath{\mathcal{O}_{#1}}\xspace}                       
\def\Opep#1   {\ensuremath{\mathcal{O}_{#1}^{'}}\xspace}                    

\newcommand{\tev}{\ensuremath{\mathrm{\,Te\kern -0.1em V}}\xspace}
\newcommand{\gev}{\ensuremath{\mathrm{\,Ge\kern -0.1em V}}\xspace}
\newcommand{\mev}{\ensuremath{\mathrm{\,Me\kern -0.1em V}}\xspace}
\newcommand{\kev}{\ensuremath{\mathrm{\,ke\kern -0.1em V}}\xspace}
\newcommand{\ev}{\ensuremath{\mathrm{\,e\kern -0.1em V}}\xspace}
\newcommand{\gevc}{\ensuremath{{\mathrm{\,Ge\kern -0.1em V\!/}c}}\xspace}
\newcommand{\mevc}{\ensuremath{{\mathrm{\,Me\kern -0.1em V\!/}c}}\xspace}
\newcommand{\gevcc}{\ensuremath{{\mathrm{\,Ge\kern -0.1em V\!/}c^2}}\xspace}
\newcommand{\gevgevcccc}{\ensuremath{{\mathrm{\,Ge\kern -0.1em V^2\!/}c^4}}\xspace}
\newcommand{\mevcc}{\ensuremath{{\mathrm{\,Me\kern -0.1em V\!/}c^2}}\xspace}

\def\m    {\ensuremath{\mathrm{ \,m}}\xspace}

\def\cm   {\ensuremath{\mathrm{ \,cm}}\xspace}

\def\mm   {\ensuremath{\mathrm{ \,mm}}\xspace}

\def\mum  {\ensuremath{{\,\upmu\mathrm{m}}}\xspace}

\def\mbarn{\ensuremath{\mathrm{ \,mb}}\xspace}

\def\sec  {\ensuremath{\mathrm{{\,s}}}\xspace}

\def\gsim{{~\raise.15em\hbox{$>$}\kern-.85em
          \lower.35em\hbox{$\sim$}~}\xspace}
\def\lsim{{~\raise.15em\hbox{$<$}\kern-.85em
          \lower.35em\hbox{$\sim$}~}\xspace}

\def\mrad{\ensuremath{\mathrm{ \,mrad}}\xspace}

\def\evtgen     {\mbox{\textsc{EvtGen}}\xspace}

\def\geant      {\mbox{\textsc{Geant4}}\xspace}

\def\pythia     {\mbox{\textsc{Pythia}}\xspace}

\def\tell1  {TELL1\xspace}
\def\ukl1   {UKL1\xspace}

\newcommand{\eg}{\mbox{\itshape e.g.}\xspace}
\newcommand{\ie}{\mbox{\itshape i.e.}\xspace}

\newcommand{\br}{\ensuremath{\mathcal{B}}\xspace}

\def\pr          {{\ensuremath{p}}\xspace}
\def\antipr          {{\ensuremath{\overline{p}}}\xspace}

\def\Lb      {{\ensuremath{\Lz^0_\bquark}}\xspace}

\def\Lc      {{\ensuremath{\Lz^+_\cquark}}\xspace}
\def\Lcbar   {{\ensuremath{\Lbar{}^-_\cquark}}\xspace}   
\def\Xicz    {{\ensuremath{\PXi^0_\cquark}}\xspace}
\def\Xicbarz    {{\ensuremath{\overline{\PXi}^0_\cquark}}\xspace}
\def\Xicp    {{\ensuremath{\PXi^+_\cquark}}\xspace}
\def\Omegacz    {{\ensuremath{\POmega^0_\cquark}}\xspace}
\def\Omegacbarz    {{\ensuremath{\overline{\POmega}^0_\cquark}}\xspace}

\def\Xibm    {{\ensuremath{\PXi^-_\bquark}}\xspace}
\def\Omegabm    {{\ensuremath{\POmega^-_\bquark}}\xspace}
\def\Sigmap    {{\ensuremath{\PSigma^+}}\xspace}
\def\Xibp    {{\ensuremath{\overline{\PXi}^+_\bquark}}\xspace}  
\def\Omegabp    {{\ensuremath{\overline{\POmega}^+_\bquark}}\xspace}   

\def\Xim    {\ensuremath{\PXi^-}\xspace}
\def\Xip    {\ensuremath{\overline{\PXi}^+}\xspace}  
\def\Omegam {\ensuremath{\POmega^-}\xspace}  
\def\Omegap {\ensuremath{\overline{\POmega}^+}\xspace} 

\newcommand{\efft}{\ensuremath{\varepsilon_{t}}\xspace}
\newcommand{\effc}{\ensuremath{\varepsilon_{c}}\xspace}
\newcommand{\effs}{\ensuremath{\varepsilon_{s}}\xspace}
\newcommand{\effl}{\ensuremath{\varepsilon_{l}}\xspace}
\newcommand{\effCH}{\ensuremath{\varepsilon_{\rm CH}}\xspace}

\newcommand{\effDF}{\ensuremath{\varepsilon_{\rm DF}}\xspace}

\newcommand{\effdet}{\ensuremath{\varepsilon_{\rm det}}\xspace}

\def\invs{\ensuremath{{\mathrm{ \,s^{-1}}}}\xspace}

\def\invcmtwo{\ensuremath{{\mathrm{ \,cm^{-2}}}}\xspace}

\def\Rq{\ensuremath{{R_{\quarkbar/\quark}}}\xspace}

\def\PV               {{\ensuremath{PV}}\xspace}

\def\SV               {{\ensuremath{SV}}\xspace}

\def\Sa               {{\ensuremath{\rm \sf S1}}\xspace}
\def\Sb              {{\ensuremath{\rm \sf S2}}\xspace}

\def\W               {{\ensuremath{\rm W}}\xspace}
\def\Si               {{\ensuremath{\rm Si}}\xspace}
\def\Ge              {{\ensuremath{\rm Ge}}\xspace}
\newcommand{\pot}{\ensuremath{\mathrm{\,PoT}}\xspace}

\def\murad  {\ensuremath{{\,\upmu\mathrm{rad}}}\xspace}

\def\evtgen     {\mbox{\textsc{EvtGen}}\xspace}

\def\geant      {\mbox{\textsc{Geant4}}\xspace}

\def\pythia     {\mbox{\textsc{Pythia}}\xspace}
\def\epos       {\mbox{\textsc{Epos}}\xspace}

\newcommand{\tevc}{\ensuremath{{\mathrm{\,Te\kern -0.1em V\!/}c}}\xspace}
\newcommand{\tevtevcccc}{\ensuremath{{\mathrm{\,Te\kern -0.1em V^2\!/}c^4}}\xspace}
\newcommand{\gevgevcc}{\ensuremath{{\mathrm{\,Ge\kern -0.1em V^2\!/}c^2}}\xspace}
\newcommand{\tevtevcc}{\ensuremath{{\mathrm{\,Te\kern -0.1em V^2\!/}c^2}}\xspace}

\newcommand{\phm}{\ensuremath{\phantom{-}}}
\newcommand{\phz}{\ensuremath{\phantom{0}}}
\newcommand{\phd}{\ensuremath{\phantom{.}}}

\begin{document}
%
\title{Electromagnetic dipole moments of charged baryons with bent crystals at the \lhc}
\author{
E.~Bagli\inst{1},
L.~Bandiera\inst{1},
G.~Cavoto\inst{2},
V.~Guidi\inst{1},
L.~Henry\inst{3},
D.~Marangotto\inst{4},
F.~Martinez Vidal\inst{3},
A.~Mazzolari\inst{1},
A.~Merli\inst{4,5},
N.~Neri\inst{4,5}, 
J.~Ruiz Vidal\inst{3}
}                     
%
\mail{
}
\institute{
       INFN Sezione di Ferrara and Universit\`a di Ferrara, Ferrara, Italy
  \and INFN Sezione di Roma and ``Sapienza'' Universit\`a di Roma, Roma, Italy
  \and IFIC, Universitat de Val\`encia-CSIC, Valencia, Spain
  \and INFN Sezione di Milano and Universit\`a di Milano, Milan, Italy
  \and CERN, Geneva, Switzerland
}
\date{Received: date / Revised version: date}
%
\abstract{
We propose a unique program of measurements of electric and magnetic dipole moments of charm, beauty
and strange charged baryons at the \lhc, based on the phenomenon of spin precession of channeled particles in bent crystals.
Studies of crystal channeling and spin precession of positively- and negatively-charged particles are presented,
along with feasibility studies and expected sensitivities for the proposed experiment
using a layout based on the \lhcb detector.

} 
\authorrunning{E.~Bagli et al.}   
\titlerunning{Electromagnetic dipole moments of charged baryons with bent crystals at the \lhc}   
\maketitle
%
%
\section{Introduction}
\label{sec:intro}
%
%
The magnetic dipole moment (MDM) and the electric dipole moment (EDM) are static 
properties of particles that determine the spin motion in an external electromagnetic field, as
 described by the T-BMT equation~\cite{Thomas:1926dy,Thomas:1927yu,Bargmann:1959gz}. 
Several measurements of baryon MDMs contributed to confirm the validity of the quark model~\cite{Olive:2016xmw}.
Measurements of the MDM of heavy baryons, \ie baryons containing charm or beauty quarks, have never been
performed due to the difficulties imposed by the short lifetime of these particles of about
$10^{-13}-10^{-12}\sec$. These measurements would provide important anchor points for QCD calculations,
helping to discriminate between different models~\cite{Sharma:2010vv,Dhir:2013nka}, 
and would improve the current understanding of the internal structure of hadrons.
The possibility to measure the MDM of positively-charged charm baryons at the Large Hadron Collider (\lhc)
using bent crystals has been proposed in Refs.~\cite{Baryshevsky:2016cul,Burmistrov:2194564} and
recently revisited~\cite{Bezshyyko:2017var,Baryshevsky:2017yhk}.
 
The EDM is the only static property of a particle
that requires the violation of parity ($P$) and time reversal ($T$) symmetries
and thus, relying on \CPT invariance, the violation of \CP symmetry.
The EDM of a baryon may arise from the structure of quarks and gluons,
and any process involving a photon and a flavour-diagonal coupling. 
In the Standard Model (SM), contributions to the EDM of baryons are highly
suppressed but can be largely enhanced in some of its extensions.
Hence, the experimental searches for the EDM of fundamental
particles provide powerful probes for 
physics beyond the SM.

Indirect bounds on charm (beauty) quark EDM are set from different experimental measurements
and span over several orders of magnitude, \ie charm (beauty) EDM $\lsim 10^{-15}-4.4\times 10^{-17}e\cm$~\cite{Sala:2013osa,Zhao:2016jcx,Grozin:2009jq,Escribano:1993xr,Blinov:2008mu} 
($\lsim 2\times 10^{-12}-2\times 10^{-17}e\cm$~\cite{Grozin:2009jq,Escribano:1993xr,Blinov:2008mu,CorderoCid:2007uc}), depending on different models and assumptions. 
As an example, an indirect bound on the
charm quark EDM is derived from the experimental limit on the neutron EDM
to be $\lsim 4.4\times 10^{-17} e\cm$~\cite{Sala:2013osa},
and a charm quark EDM of comparable magnitude is possible in extensions
of the SM~\cite{Zhao:2016jcx}. For the beauty quark, indirect EDM limits
$\lsim 2\times 10^{-12} e\cm$~\cite{Grozin:2009jq} and $\lsim 1.22\times 10^{-13} e\cm$~\cite{CorderoCid:2007uc}
are derived, and a relatively large beauty quark EDM is possible in presence of 
new physics.

Recently, it has been proposed to
search for the EDM of positively-charged charm baryons using bent crystals at the \lhc~\cite{Botella:2016ksl}.
Similarly to the MDM case, the method relies on baryons produced by the interaction of $7 \tev$
protons, extracted from the \lhc beam, on a fixed target. The baryons are subsequently channeled in a
bent crystal. The spin precession of short-lived particles is induced 
by the intense electromagnetic field between the crystal atomic planes. The
EDM and the MDM information can be extracted by measuring the spin polarization of the channeled
baryons the end of the bent crystal.
This technique can be extended to strange
and beauty positively-charged baryons.
%
%
%
%

In this paper we address several key aspects 
of this unique experimental program.
In Secs.~\ref{sec:channeling} and~\ref{sec:spin_geant4}, 
after introducing the channeling of charged particles in bent crystals,
we study the deflection and the spin precession of positively- and
negatively-charged baryons in a bent crystal using \geant simulations.
We assess the experimental technique and study the possibility to 
extend the EDM searches and MDM measurements to negatively-charged baryons.
This would allow to perform tests of the \CPT symmetry by measuring the MDM of particles and antiparticles. 
In Sec.~\ref{sec:spin_precession} we prove that the spin evolution equations describing MDM and EDM effects hold for non-harmonic 
planar channel potential, therefore spin precession effects for both positively- and negatively-charged particles depend 
uniquely on the crystal curvature. In fact, the same equations also apply for axial-channeled particles, 
mostly relevant for negatively-charged particles, although its application to MDM and EDM physics will require further investigation.
Section~\ref{sec:setup} focuses on the description of a possible fixed-target setup installed 
in front of the \lhcb detector.
The feasibility of the measurements has been evaluated relying on both parametric and \geant simulations 
along with a geometrical model of the detector. 
Finally, Sec.~\ref{sec:sensitivity} presents sensitivity studies for EDM searches and MDM measurements of 
$\Lc$, $\Xicp$ charm baryons, 
$\Xibp$, $\Omegabp$ beauty antibaryons, and
$\Xip$, $\Omegap$ strange antibaryons.
Baryons and antibaryons will be referred hereafter generically as baryons, unless otherwise stated.

\section{Channeling of multi-TeV charged particles}
\label{sec:channeling}

In a crystal the strong electric field experienced by a charged particle in the proximity of the ordered structure of the atoms exerts a strong confinement force 
onto the particle itself. The particle trajectory  can be bound   to stay parallel to
a crystalline plane or to an atomic string, which becomes a preferential pathway in the crystal. This phenomenon is
called {\it channeling} and can occur if the angle between the particle trajectory and a crystal plane ({\it planar }
channeling)  or  a crystal axis ({\it axial} channeling)  is lower than a Lindhard angle
$\theta_L = \sqrt{2U_0/(p\beta c)}$, where $U_0$ is the potential-well depth, $p$ the particle momentum and $\beta$ its velocity~\cite{Lindhard,Biryukov1997}. 
This process has been studied  in laboratory up to the highest available  energy, in particular at the LHC where the planar channeling of 6.5\tev protons has been observed \cite{Scandale:2016krl}.

When the crystal is bent, its planes or atomic strings are bent too. 
The particle pathways are adiabatically bent following the crystal curvature, resulting in a net deflection
of the incoming direction by an angle equal to that of crystal bending:
charged particle steering is then possible through channeling in bent crystals. 
Various applications as circular accelerator halo collimation
\cite{Scandale:2015gva,Scandale:2013lea,Scandale:2012wy,Scandale:2011za,Scandale:2010zzb} or beam extraction from
an accelerator ring for fixed-target experiments \cite{Lansberg:2015} have been studied and proposed also for the \lhc.

Beam steering of positively-charged particles (positive particles in the following) with channeling   has progressed significantly over the last  years, featuring silicon  crystals  with about 80\%   deflection efficiency  at an energy of several hundred GeV \cite{Scandale:2008zzk,Scandale:2009zzd}. Positive particles in channeling condition  are repelled by the atomic electric field and follow trajectories that tend to be far from the lattice sites.  Negative particles, on the contrary, are attracted by the same field and can repeatedly oscillate across the nuclei of the crystal. For this reason negative  particles are more likely to collide   with the nuclei of the crystal lattice  and therefore can easily escape from a channeling bound state.  The average length that a channeled particle traverses before exiting from the planar or axial potential well is called
{\it dechanneling} length $L_d$,
being much shorter for negative than for positive charges.

The first  measurements of  $L_d$ for ultra-high energy negative particles has been done at CERN using relatively short bent silicon crystal at few hundreds \gev. It has been measured to be few \mm \cite{Scandale:2009zzb,Scandale:2013rva,Bagli2017}
using negative hadron and electron beams,
and successfully  compared with simulations. Therefore, simulations can  be used to extrapolate the efficiency of channeling processes to the multi-TeV energy range  and for different crystal curvature radii.

%
%
%
%
%
%

For the  ultra-relativistic energy range it was also demonstrated the ability to steer  negative particle beams through the axial channeling regime with an  efficiency above 90\%~\cite{Scandale:2009zze,Scandale:2010zze}. Indeed, in the case of axial channeling,  there are particles which are above the electric field barrier and are not trapped along a single  atomic string. Those particles   are anyway deflected due to their stochastic scattering with different  atomic strings of the crystal, avoiding the fast dechanneling occurring for negative particles. However, reaching a condition of axial alignment for beam steering is relatively more difficult than for planar channeling since the orientation of the crystal with respect to two (and not only one) rotational axes must be found.

 The dependence of the channeling efficiency for positive particles on the particle energy, the crystal length, and bending radius is well known for crystals with the length along the beam comparable to $L_{d}$ \cite{Biryukov1997}.
 The dechanneling length scales almost proportionally to the particle
 momentum-velocity $p{\beta}$, \ie $L_{d} \propto p{\beta}$,
 and can be calculated as
\begin{equation}
L_{d} = \frac{256}{9\pi^2} \frac{p\beta c}{\ln(2m_e c^2 \gamma/I)-1} \frac{a_{\rm TF}d_p}{Z_i r_e m_e c^2} ,
\end{equation}
where $m_e$ and $r_e$ are the mass and the classical radius of the electron, $I$ is the mean ionization energy, $d_p$ is the interplanar spacing, $a_{\rm TF}$ is the Thomas-Fermi screening radius, and $Z_i$ and $\gamma$  are the charge number and boost of the incident particle.

The channeling efficiency under harmonic approximation scales proportionally to $1 - \eta_{c}$, with $\eta_{c} = R_{c}/R$, where $R$ 
is the crystal bending radius and $R_{c} \propto p{\beta}$ is the critical radius for channeling, \ie the minimum bending radius for which channeling occurs, and holds,
\begin{equation}
\epsilon(R) = \epsilon(R=\infty)\left(1-\eta_{c}\right).
\end{equation}
Since the fraction of particles which remains channeled in a bent crystal has to oscillate
near the potential well edge, the probability to leave the channeling state increases
for small bending radii.
Under harmonic approximation for a bent crystal the dechanneling length is shortened by a factor $(1-\eta_{c})$, 
\begin{equation}
L_{d}(R)=L_{d}(R=\infty)\left(1-\eta_{c}\right).
\end{equation}

Differently, the energy dependence of $L_{d}$ for negative particles is not theoretically  known.
Contrarily to the dechanneling of positive particles, the large variation of the transverse energy by
negative particles in the interaction with atomic nuclei cannot be treated as a stochastic process,
and requires a new formalism to be developed.

To quantify the deflection efficiency of the channeling process in bent crystals, Monte Carlo simulations for both positive and negative particles have been carried out using the \geant toolkit~\cite{Agostinelli2003250,Allison2016186} in a version allowing for crystalline structures~\cite{Bagli2017SolidStructure}. The channeling process is implemented by including \textsc{Dynecharm++}~\cite{Bagli2013124} and \textsc{Echarm}~\cite{PhysRevE.81.026708} into the \geant channeling package~\cite{s10052-014-2996-y}. Such model was validated against experimental data for negative pions \cite{Scandale:2013rva}, protons \cite{PhysRevLett.110.175502,SCANDALE20141,bagli201474,bagli2014743114,BAGLI2015387,PhysRevLett.115.015503,SCANDALE2016826}, electrons \cite{PhysRevLett.114.074801,PhysRevLett.115.025504,BANDIERA201544,PhysRevAccelBeams.19.071001,Bagli2017} and positrons \cite{Bagli2017} in a range of energies spanning from $855 \mev$ (electrons at MAMI) to $400 \gev$ (protons at CERN). \textsc{Dynecharm++} allows the tracking of a relativistic charged particle inside a crystalline medium via the numerical integration of the T-BMT classical equations of motion~\cite{Thomas:1926dy,Thomas:1927yu,Bargmann:1959gz}. 
The continuum potential approximation proposed by Lindhard is used \cite{Lindhard}. \textsc{Echarm} allows the computation if the electrical characteristics of the crystal is within this approximation.

The \geant application was developed on top of the 10.3 version of the toolkit, which allows for crystalline structures \cite{Bagli2017SolidStructure}. 
The \geant physics lists used were the \texttt{G4HadronElasticPhysics} and the \texttt{G4HadronPhysicsFTFP\_BERT} and a custom 
\texttt{G4EmStandardPhysics\_option4} with single scattering instead of multiple-scattering.
\begin{figure}[htb!]
\includegraphics[width=1.0\columnwidth]{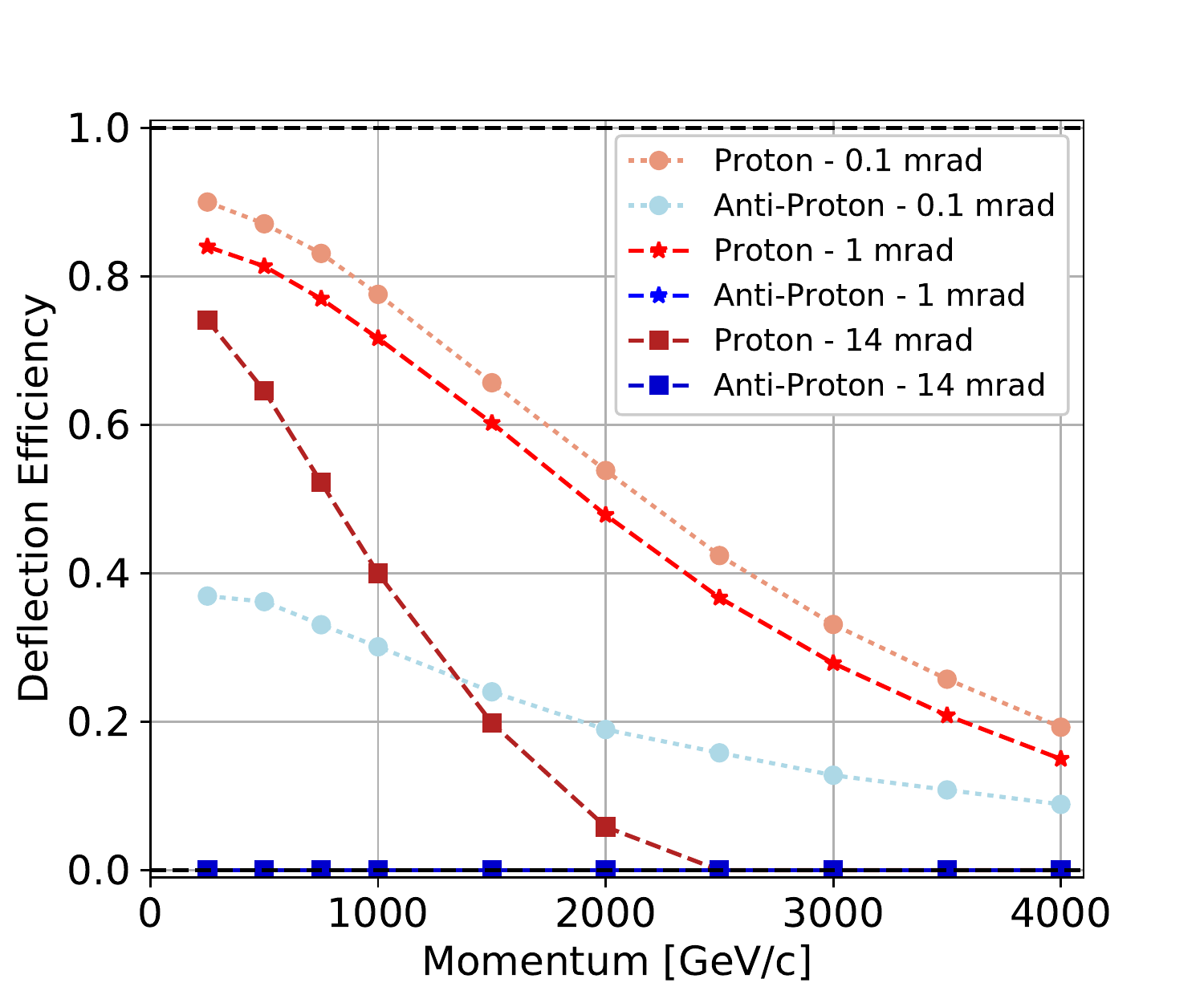}
\caption{Dependence of the channeling efficiency of protons and antiprotons with the particle momentum for 1 \mm, 1\cm and 7\cm long \Si crystals 
bent along the $(110)$ plane by a $0.1$\mrad, $1$\mrad and $14$\mrad bending angle, respectively. The curves for the anti-proton interacting with the  1\cm and 7\cm long \Si crystals are superimposed. 
}
\label{fig:geant4_efficiency_dep}
\end{figure}

\begin{figure}[htb!]
\includegraphics[width=0.9\columnwidth]{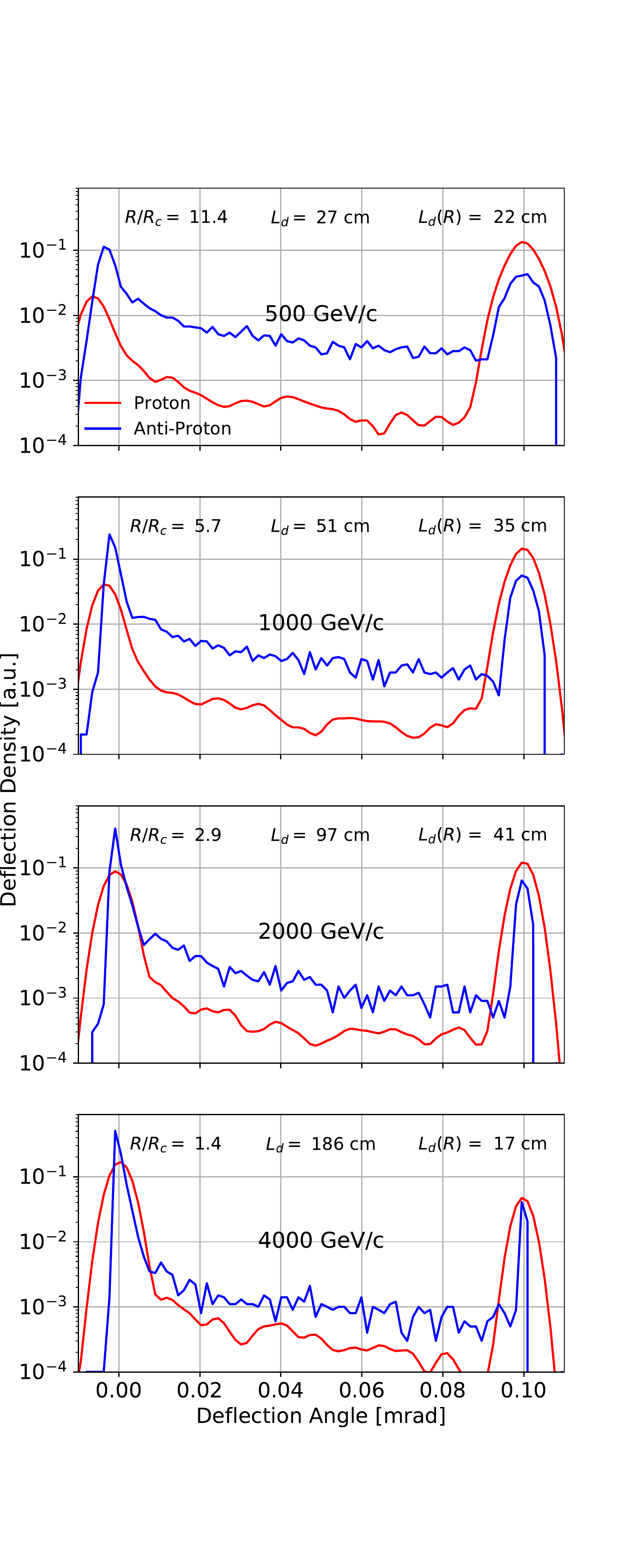}
\caption{Outgoing angular distributions at various momenta for protons and antiprotons impinging on a $1$\mm long \Si crystal bent along the $(110)$ plane by a $0.1$\mrad bending angle. 
The dechanneling length for positive particles in straight crystals ($L_{d}$) 
and in the bent crystal ($L_{d}(R)$), along with the critical radius for channeling ($R_{c}$),
are calculated for the different momenta following Ref.~\cite{Biryukov1997}.}
\label{fig:geant4_efficiency_def}
\end{figure}

Figure~\ref{fig:geant4_efficiency_dep} shows the dependence of the channeling efficiency for protons and antiprotons with the particle momentum for 
a $1$\mm, $1$\cm, $7$\cm long \Si crystal bent along the $(110)$ plane by a
$0.1$\mrad, $1$ \mrad  and $14$ \mrad bending angle, respectively.
The efficiency for positive particles is not spoiled by $L_{d}$, 
being  $L_{d}>13$ cm for all the momenta, but   the unfavourable  ratio between the critical radius $R_{c}$ and the bending radius is causing it to 
decrease. Indeed, as illustrated in Fig.~\ref{fig:geant4_efficiency_def}, this ratio rapidly increases, lowering the deflection efficiency. 
For negative particles, the deflection efficiency is largely dominated
by the crystal length. Therefore, the efficiency remains always lower than for positive particles for all the momenta.
Such simulations show that the crystal geometric parameters have to be
carefully  chosen depending on the energy range in which the crystal has
to be operated. In the figure, particles which are not captured under channeling at the crystal entrance are reflected to the opposite side 
with an angle which depends on the particle momentum \cite{PhysRevLett.98.154801}.

\section{Spin precession}
\label{sec:spin_precession}



The spin precession of a charged particle is induced by the interaction of its electromagnetic
dipole moments, \eg MDM and EDM, with external electromagnetic fields. 
The time evolution of the spin-polarization vector $\mathbf{s}$ is regulated by the T-BMT equation
\begin{equation}
\frac{d\mathbf s}{dt}=\mathbf s\times \mathbf\Omega, \hspace{0.4cm} \mathbf\Omega=\mathbf\Omega_{\rm MDM}+\mathbf\Omega_{\rm EDM}+\mathbf\Omega_{\rm TH},
\label{eq:spin_precession}
\end{equation}
where the precession angular velocity vector $\mathbf\Omega$ is composed by three contributions corresponding to the
MDM, EDM, and Thomas precession:
\begin{eqnarray}
  \mathbf\Omega_{\rm MDM} &=& \frac{g \mu_B}{\hslash}\left(\mathbf B -\frac{\gamma}{\gamma+1}({\bm \beta}\cdot\mathbf B){\bm \beta}-{\bm\beta}\times\mathbf E\right),\nonumber\\
  \mathbf\Omega_{\rm EDM} &=& \frac{d \mu_B}{\hslash}\left(\mathbf E -\frac{\gamma}{\gamma+1}({\bm \beta}\cdot\mathbf E){\bm \beta}-{\bm\beta}\times\mathbf B\right),\nonumber\\
 \mathbf\Omega_{\rm TH}\hspace{0.25cm} &=& \frac{\gamma^2}{\gamma+1}\bm\beta\times\frac{d\bm\beta}{dt}=   
 \frac{q}{mc}\bigg[\left(\frac{1}{\gamma}-1\right)\mathbf B+ \nonumber\\
&&   \frac{\gamma}{\gamma+1}(\bm \beta \cdot \mathbf B)\bm \beta -
   \left(\frac{1}{\gamma+1}-1\right)\bm \beta \times \mathbf E\bigg],\nonumber\\
 \label{eq:omega_contributions}
\end{eqnarray}
where $\mathbf E$ and $\mathbf B$ are the electric and the magnetic fields in the laboratory frame, 
and $q$, $\gamma$ and $\bm \beta$ are the electric charge, boost and vector velocity of the particle, respectively.
The $g$ and $d$ dimensionless factors, also referred to as the gyromagnetic and gyroelectric ratios,
define the magnetic and electric dipole moment of a 
particle with spin $J$ (in Gaussian units) as $\bm{\mu} = J g \mu_B {\mathbf s}$ and \mbox{$\bm{ \delta} = J d \mu_B {\mathbf s}$}, respectively,
where $\mu_B=e \hbar / (2 m c)$ is the particle magneton\footnote{The spin-polarization vector is defined such as 
${\mathbf s} = \langle {\mathbf S} \rangle / (J \hbar)$, where ${\mathbf S}$ is the spin operator.}.

The lifetime of baryons with heavy quark constituents is too short, \eg the \Lc baryon lifetime is about
$10^{-13}\sec$, for a standard magnet to induce any detectable effect to the spin-polarization vector before they decay. 
The possibility to measure the MDM of short-lived baryons using channeling in bent crystals was firstly pointed out 
by V.G. Baryshevsky in 1979.
The method is based on the interaction of the MDM of the channeled particles
with the intense electric field between crystal atomic planes.
As an example, a sketch of the deflection of the \Lc baryon trajectory and spin precession in a bent crystal
is shown  in Fig.~\ref{fig:CrystalPlane}.
 Charm baryons produced by interaction of protons on a fixed target, \eg a \W target, are polarized
 perpendicularly to the production plane due to parity conservation in strong interactions.
The production plane $xz$, also shown in the figure, is determined by the proton and
 baryon momenta; the latter defines the $z$ axis.
 The initial polarization vector $\mathbf{s_0} = (0, s_0, 0)$ is perpendicular to the production plane,
 along the $y$ axis, and at the end of the crystal it is rotated of an angle $\Phi$.
The crystal is bent in the $yz$ plane by an angle $\theta_C$.
%
The measurement of the MDM of charm baryons using bent crystals has been widely discussed since the 80's~\cite{Kim:1982ry,Baublis:1994ku,Samsonov:1996ah}. 
Lately, the possibility of the measurement at \lhc energies has been considered~\cite{Baryshevsky:2016cul,Burmistrov:2194564,Bezshyyko:2017var}. 
Recently, the search for charm baryon EDM using bent crystals at \lhc has been proposed~\cite{Botella:2016ksl}. 

The spin precession of particles channeled in bent crystals was firstly observed by the E761
Collaboration~\cite{Chen:1992wx}. 
 Using a 800 \gevc proton beam impinging on a Cu target, 
$\Sigma^+$ baryons with 375 \gevc average momentum were produced and channeled in two bent crystals with opposite bending
 angles.
The MDM of the $\Sigma^+$ baryon was measured and proved the viability of this technique for
the measurement of the MDM of short-lived particles. 

\subsection{Planar channeling}
\label{sec:spin_precession_planar}

In the case of planar channeling,
the intense electric field between the crystal planes, $\mathbf{E}$, which deflects charged particles,
transforms into a strong electromagnetic field 
$\mathbf{E^*} \approx \gamma \mathbf{E}$,
$\mathbf{B^*} \approx - \gamma\boldsymbol{\beta} \times \mathbf{E}/c$ 
in the instantaneous
rest frame of the particle and induces spin precession. In the limit of large boost, the spin precession
induced by the MDM in the $yz$ plane is~\cite{Lyuboshits:1979qw} 
\begin{equation}
\varPhi \approx \frac{g-2}{2}\gamma\theta_C .
\label{eq:Phi}
\end{equation}
In order to obtain $\Phi\approx 1$ rad are required large crystal bending angles $\theta_C\approx 1-10\mrad$,
and large Lorentz factors $\gamma\approx 10^2-10^3$, which can be uniquely achieved at \lhc.
\begin{figure}[htb]
\includegraphics[width=1\columnwidth]{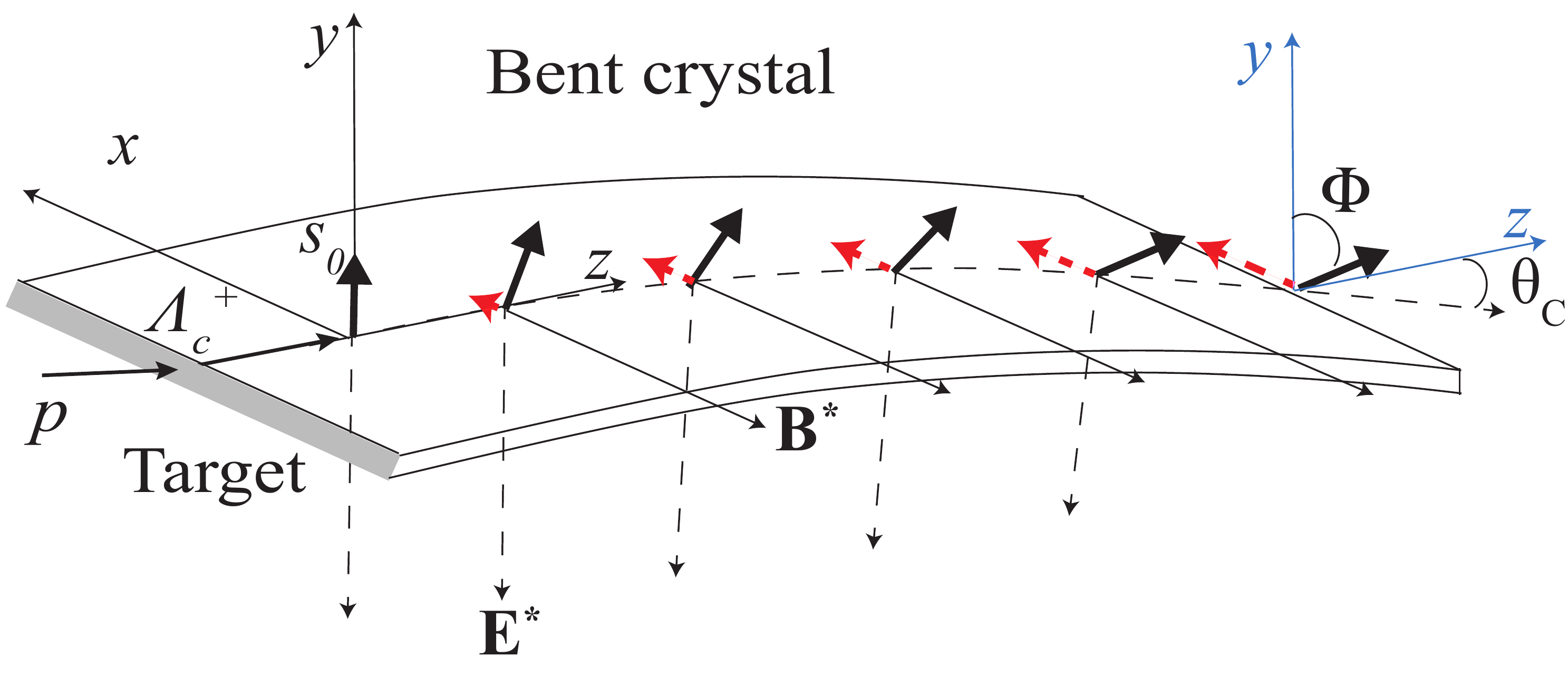}
\caption{Sketch of the deflection of the \Lc baryon trajectory and spin precession in a bent crystal.
  The initial polarization vector $\mathbf{s_0}$ is perpendicular to the production plane,
  along the $y$ axis, due to parity conservation in strong interactions.
  The spin precession in the $yz$ and $xy$ plane are induced by the MDM and the EDM, respectively.
  The red (dashed) arrows indicate the (magnified) $s_x$ spin-polarization component proportional to the particle EDM.
  The $\Phi$ angle indicates the spin precession due to the MDM. 
}
\label{fig:CrystalPlane}
\end{figure}

\begin{figure}[htb]
\includegraphics[width=1\columnwidth]{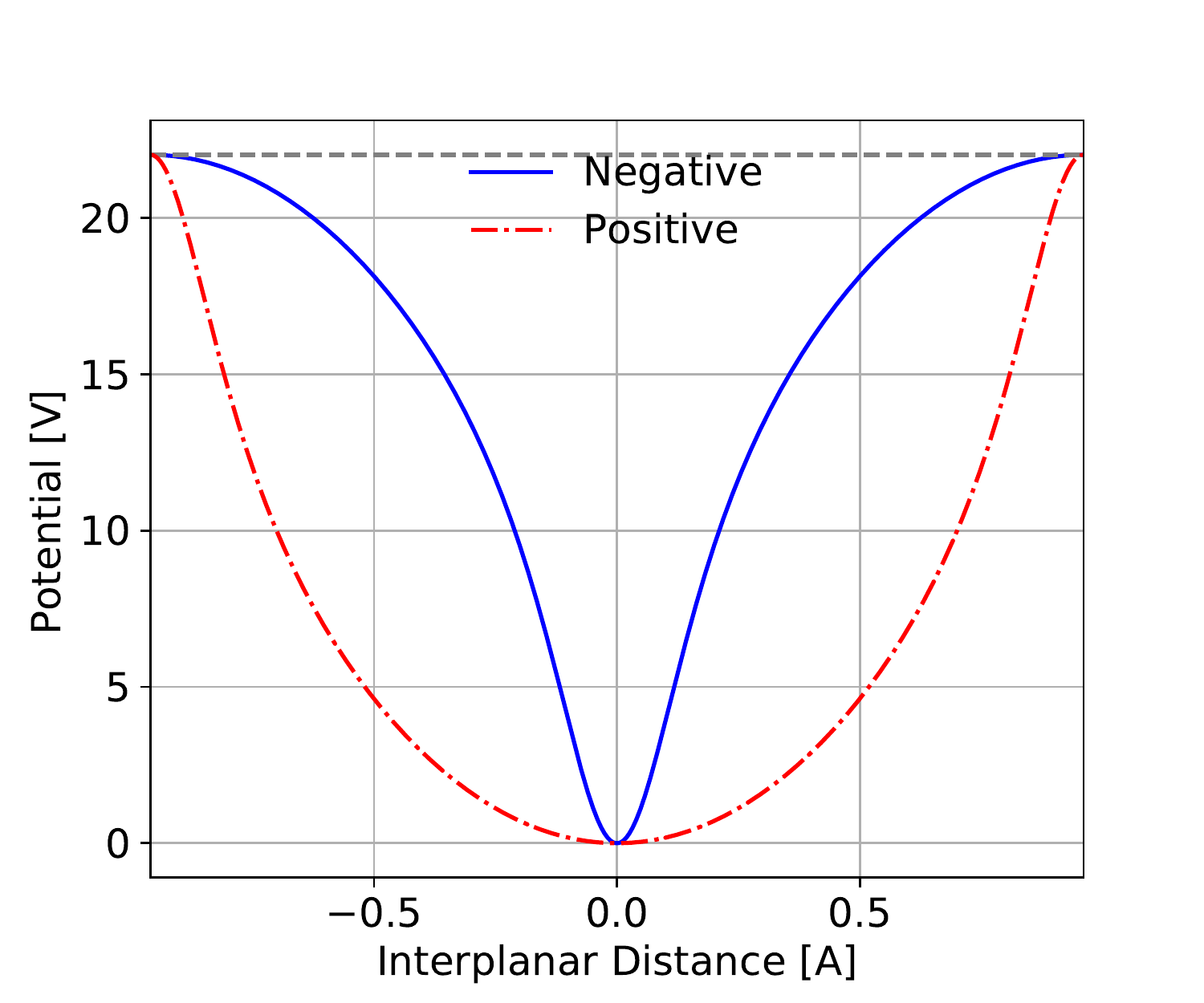}
\caption{Harmonic (red dash-dotted line) and non-harmonic (blue continuous line) electric potential versus interplanar distance for positive 
and negative particles in a $(110)$ \Si crystal.
The electric potential is extracted from \geant simulations. For the sake of comparison the electric
potential for negative particles is  shifted by half of the interplanar distance.
}
\label{fig:potential_negativeP}
\end{figure}

The equations describing the spin precession of planar channeled positive particles in presence
of MDM and EDM are derived in Ref.~\cite{Botella:2016ksl}. In the limit of large boost, and assuming small EDM effects compared to the 
main MDM spin precession, a polarization component orthogonal to the bending plane is induced,
\begin{equation}
s_{x}   \approx  s_0 \dfrac{d}{g-2}  (\cos{\Phi}-1).
\label{eq:EDM_precession}
\end{equation}
%
%
The MDM driven precession taking place in the bending plane is given by
\begin{equation}
\begin{aligned}
s_y  & \approx s_0 \cos\varPhi , \\
s_z  & \approx s_0 \sin\varPhi .
\end{aligned}
\label{eq:MDM_precession}
\end{equation}
Inside the crystal, positive particles feel a non-zero mean electric field thanks to the centripetal force induced by the crystal bending, and the spin precession depends basically on $\theta_C$.
The planar channel potential seen by positive particles can be assumed to be approximately harmonic, as described in Sec.~\ref{sec:channeling}. For negative particles this assumption is no more valid since their motion is regulated
by a non-harmonic potential, as shown in Fig.~\ref{fig:potential_negativeP}.

In the following, we demonstrate that in presence of a non-harmonic potential $V$, identical spin precession equations derived for the harmonic potential case
hold.
We consider the layout of Fig.~\ref{fig:CrystalPlane}, with the crystal bent along an atomic plane. Polar coordinates are introduced for describing the particle trajectory in the bending plane
\begin{equation}
y(t) = \rho(t)\cos(\Omega t), \hspace{0.4cm} z(t) = \rho(t)\sin(\Omega t),
\label{eq:radial_coordinates_planar}
\end{equation}
where $\Omega$ is the revolution frequency for the particle traversing the bent crystal, 
and the electric field described by the planar channel potential $V(\rho)$ is
\begin{equation}
\mathbf E ~=~
\left\{
\begin{aligned}
E_x &= 0\\
E_y &= -\frac{dV}{d\rho} \cos(\Omega t)\\
E_z &= -\frac{dV}{d\rho} \sin(\Omega t)~.
\end{aligned}
\right.
\label{eq:electric_field_planar}
\end{equation}
Neglecting EDM contributions, the spin evolution resulting from Eqs.~\eqref{eq:spin_precession},~\eqref{eq:omega_contributions},~\eqref{eq:radial_coordinates_planar} and \eqref{eq:electric_field_planar} is
\begin{equation}
\mathbf s(t) ~=~
\left\{
\begin{aligned}
s_x(t) &= 0\\
s_y(t) &= s_0\cos\left( \frac{2\mu'\Omega}{\hbar c} \int^t_0 \rho \frac{dV}{d\rho} dt' \right)\\
s_z(t) &= s_0\sin\left( \frac{2\mu'\Omega}{\hbar c} \int^t_0 \rho \frac{dV}{d\rho} dt' \right),
\end{aligned}
\right.
\label{eq:spin_precession_noedm}
\end{equation}
for the initial condition ${\mathbf s_0} = \left( 0, s_0, 0\right)$ and where 
\begin{equation}
\mu'\equiv\frac{g-2}{2}\frac{e\hbar}{2mc}.
\end{equation}
The radial coordinate $\rho$ is constant up to $\delta\rho/\rho = \mathcal{O}(\mathrm{\AA}/m) = 10^{-10}$, therefore the spin precession depends on 
$\int^t_0 dV/d\rho~dt'$. Over a complete oscillation in the channel potential the effect of this term is 
equivalent to that of the electric field in the particle
equilibrium radial position $\rho'_0$,
\begin{equation}
\int^t_0 \frac{dV}{d\rho} dt'=-E(\rho'_0) t,
\label{eq:equivalence}
\end{equation}
which is determined solely by the centripetal force $f_c$ induced by the bent
 trajectory,
\begin{equation}
E(\rho'_0) = -\frac{f_c}{e} = -\frac{m\gamma c^2}{e\rho'_0}.
\label{eq:centrifugal_force}
\end{equation}
This statement follows by computing
\begin{equation}
\int^t_0  \frac{dV}{d\rho} dt' -\left[-E(\rho_0')t\right] = \int^t_0 \left( \frac{dV}{d\rho} -\frac{f_c}{e} \right) dt'
\end{equation}
for a complete particle oscillation. By changing the integration variable to $d\rho$ and $dt'=d\rho/\dot{\rho}$,
then $\dot{\rho}$ is determined by the non-relativistic energy conservation for the radial motion of channeled particles~\cite{Biryukov1997}
\begin{equation}
\frac{1}{2} M\dot{\rho}^2 +e V(\rho) -f_c\rho = W_{r},
\end{equation}
in which $M=m\gamma$ and $W_{r}$ is the total radial energy, assumed to be constant during a particle oscillation. The relation holds because the longitudinal motion is ultra-relativistic and independent from the radial one, which is non-relativistic since the potential depth is $\mathcal{O}(100 \ev)\ll m$. The integration boundaries $\rho_{1,2}$ are chosen to be the particle oscillation limits, in which
\begin{equation}
\frac{1}{2} M\dot{\rho}^2 =0 \leftrightarrow e V(\rho_{1,2}) -f_c\rho_{1,2} = W_{r}~.
\end{equation}
Finally, the integral can be trivially computed
\begin{align}
&\sqrt{\frac{m}{2e}} \int^{\rho_2}_{\rho_1}  \frac{e\frac{dV}{d\rho} - f_c}{\sqrt{W_{r} +f_c\rho -e V(\rho)}} d\rho \nonumber\\
&= -\sqrt{\frac{m}{2e}} \left( \sqrt{W_{r} +f_c\rho_2 -e V(\rho_2)}\right. \nonumber\\
&\left. - \sqrt{W_{r} +f_c\rho_1 -e V(\rho_1)} \right)\nonumber\\
&=0.
\end{align}
Summarizing, spin precession effects given by the actual shape of the planar channel potential cancel out at each
particle oscillation and the net spin precession depends uniquely on the crystal curvature.
This result generalises the same conclusion previously obtained for harmonic potentials~\cite{Kim:1982ry,Lyuboshits:1979qw}.
The spin evolution equations describing MDM and EDM effects, Eqs.~(\ref{eq:Phi}),~(\ref{eq:EDM_precession}) and~(\ref{eq:MDM_precession}), 
hold as for an harmonic planar channel potential, and in particular for the potential seen by negative particles.

\subsection{Axial channeling}
\label{sec:spin_precession_axial}

The planar channeling efficiency for negative particles is smaller than for positive.
It becomes negligibly small for crystals longer than $1\cm$ with bending angle larger than $1\mrad$, as shown in Fig.~\ref{fig:geant4_efficiency_dep}.
In this case, the spin depolarization effect of particles scattered by crystal axes (planes) provides a
possibility for measuring the MDM of negative particles~\cite{Baryshevsky:2017yhk}.
In case of axial alignment, spin rotation can also be investigated.
The phenomenon of axial channeling has been observed for positive and negative particles but it has
not been considered for spin precession to date. Here we discuss the possibility to induce spin precession in axial-channeled particles for potential
applications in MDM and EDM measurements of charged baryons.
We consider the same layout of Fig.~\ref{fig:CrystalPlane}, in which the crystal is now bent along a crystallographic axis.
Polar coordinates are introduced for the bending plane as in Eq.~\eqref{eq:radial_coordinates_planar}
and the electric field described by the axial channel potential $V(x,\rho)$ is
\begin{equation}
\mathbf E ~=~
\left\{
\begin{aligned}
E_x &= -\frac{dV}{dx}\\
E_y &= -\frac{dV}{d\rho} \cos(\Omega t)\\
E_z &= -\frac{dV}{d\rho} \sin(\Omega t)~.
\end{aligned}
\right.
\label{eq:electric_field_axial}
\end{equation}
In analogy with planar-channeled particles described in Sec.~\ref{sec:spin_precession_planar}, the longitudinal velocity is ultra-relativistic. The velocity components orthogonal to the channel are non-relativistic; their contribution to the
spin precession described in Eq.~(\ref{eq:omega_contributions}), is negligible.
The particle velocity inside the bent crystal is therefore simplified as
\begin{equation}
\beta_x = 0, \hspace{0.4cm} \beta_y = -\sin(\Omega t), \hspace{0.4cm} \beta_z = \cos(\Omega t),
\label{eq:velocity_axial}
\end{equation}
in which $\cos(\Omega t)\approx 1$ and $\sin(\Omega t)\approx 0$ to a very good approximation for a crystal bending angle $\theta_C \leq 15$ mrad.
The spin evolves according to the spin precession equation
\begin{equation}
\frac{d\mathbf s}{dt}=\mathbf s\times \mathbf\Omega =
\left\{
\begin{aligned}
\frac{ds_x}{dt} & =  s_y\Omega_z - s_z\Omega_y\\
\frac{ds_y}{dt} & =  s_z\Omega_x - s_x\Omega_z\\
\frac{ds_z}{dt} & =  s_x\Omega_y - s_y\Omega_x~,
\end{aligned}
\right.
\label{eq:spin_precession_components}
\end{equation}
with precession vector $\mathbf \Omega$ following from Eqs.~(\ref{eq:omega_contributions}),~(\ref{eq:electric_field_axial}) and~(\ref{eq:velocity_axial}),
\begin{equation}
\mathbf \Omega ~=~
\left\{
\begin{aligned}
\Omega_x &= -\frac{2\mu'}{\hbar} \frac{dV}{d\rho} - \frac{d\mu_B}{\hbar} \frac{dV}{dx}\\
\Omega_y &= \frac{2\mu'}{\hbar} \frac{dV}{dx} - \frac{d\mu_B}{\hbar} \frac{dV}{d\rho}\\
\Omega_z &= 0~.
\end{aligned}
\right.
\label{eq:precession_vector_axial}
\end{equation}
%
The main difference with respect to the planar channeling case is the presence of the $E_x$ electric field component, which in principle complicates the separation between the MDM and EDM induced spin rotation. Nonetheless, in the following it is shown that the contribution of the $dV/dx$ terms can be neglected and the spin precession evolution derived for the planar channeling case applies also to axial-channeled particles.

During a particle oscillation the spin can be assumed to be constant since the typical spin precession frequency $\omega=2\mu'E(\rho'_0)/\hbar\approx 10^{10}$ Hz is three orders of magnitude lower than the oscillation frequency of the particle trapped in the channel, $\Omega_k \approx 10^{13}$ Hz.
The two dominant components $\Omega_x \propto dV/d\rho$ and $\Omega_y \propto dV/dx$, describing spin precession in the $yz$ and $xz$ planes, respectively, can be considered to act independently of each other; namely, the spin rotation in the $yz$ plane is not influenced by the spin rotation in the $xz$ plane and vice versa. In this case Eq.~(\ref{eq:equivalence}) can be applied to both contributions: while the centripetal force induces a net spin precession in the $yz$ plane identical to that of planarly-channeled particles, the effect of $dV/dx$ mediates to zero over each particle oscillation, since no centripetal force acts in the $x$ direction.

The limit of the employed assumption is checked estimating the typical amount of spin precession accumulated during an incomplete particle oscillation, which may lead to an imperfect cancellation of the $dV/dx$ contribution. This amount is at the order of
\begin{equation}
\Delta \approx \frac{2|\mu'|}{\hbar} \int_{\rm half} |\mathbf E| dt \approx \frac{2|\mu'|}{\hbar} \frac{|\mathbf E|}{2\Omega_k} \approx 1.5 \times 10^{-4},
\end{equation}
in which the integration is carried on half of an oscillation.
Here, $\mu'$ is taken with $(g-2)/2 = -0.3$ and the \Lc mass~\cite{Olive:2016xmw}.
The typical electric field magnitude of the axial channel $|\mathbf E|\approx 4 \times 10^{11}\ev/\m$ is estimated as the ratio between the potential depth $\approx 200 \ev$ and the channel width $\approx 5 \mathrm{\AA}$ for a Ge crystal, with values taken from Ref.~\cite{Biryukov1997}. The oscillation frequency is $\Omega_k = \sqrt{kc^2/eW} \approx 5.4 \times 10^{13}$ for $W = 1\tev$, where the constant describing the potential curvature $k\approx 3.2 \times 10^{22} \ev/\m^2$ for a Ge axial channel is about eight times the value for a \Si axial channel, according to Ref.~\cite{Biryukov1997}.

Neglecting EDM effects ($d=0$), Eqs.~\eqref{eq:spin_precession_components} and ~\eqref{eq:precession_vector_axial} show that $dV/dx$ contributes to the spin component $s_x$ via $ds_x/dt = - s_z\Omega_y$.
Since the $s_z$ spin component is not constant during a complete oscillation, the $dV/dx$ contribution is not
exactly zero and can be conservatively estimated in Eq.~(\ref{eq:deltasz}),
in which $s_z(t)=\overline{s}_z+\delta s_z(t)$ changes by an amount of order $\Delta$,
\begin{align}
 \label{eq:deltasz}
\delta s_x &\approx \frac{2\mu'}{\hbar} \int s_z(t) \frac{dV(t)}{dx} dt \approx \Delta \frac{2\mu'}{\hbar} \int_{\rm half} \frac{dV(t)}{dx} dt \nonumber\\
&\approx \Delta^2 \approx 2 \times 10^{-8}.
\end{align}
%
The integrated effect along the whole crystal is conservatively estimated by multiplying $\delta s_x$ by the number of particle oscillations,
\begin{equation}
\Delta s_x \approx \delta s_x \frac{L\Omega_k}{c} \approx 3.5 \times 10^{-4},
\end{equation}
in which a crystal length of $L=10\cm$ is taken. Such a component does not affect the main MDM spin precession in the $yz$ plane and it is negligible compared to the experimental sensitivity on the particle polarization. Indeed, according to the sensitivity studies detailed in Sec.~\ref{sec:sensitivity}, the uncertainty on the $s_x$ spin-polarization component, constituting the EDM signature, will be at the order of $10^{-2}$.

In summary, the spin evolution equations describing MDM and EDM effects, Eqs.~(\ref{eq:Phi}),~(\ref{eq:EDM_precession}) and~(\ref{eq:MDM_precession}),
hold also for axial-channeled particles, which is mostly relevant for
negative particles with relatively high axial channeling efficiency.
Nevertheless, the application of this result to the measurement of the MDM and EDM of particles has to be further studied.

%

\section{\geant simulations of spin precession}
\label{sec:spin_geant4}



The simulation method based on the numerical integration of the classical equations of motion allows to introduce
the modification of the particle spin under the effect of the strong electric field generated by the crystalline lattice.
Indeed, the step-by-step variation of the spin is tracked in \geant by numeric integration of the
T-BMT equation~\cite{Thomas:1926dy,Thomas:1927yu,Bargmann:1959gz}.

\begin{figure}
\includegraphics[width=1\columnwidth]{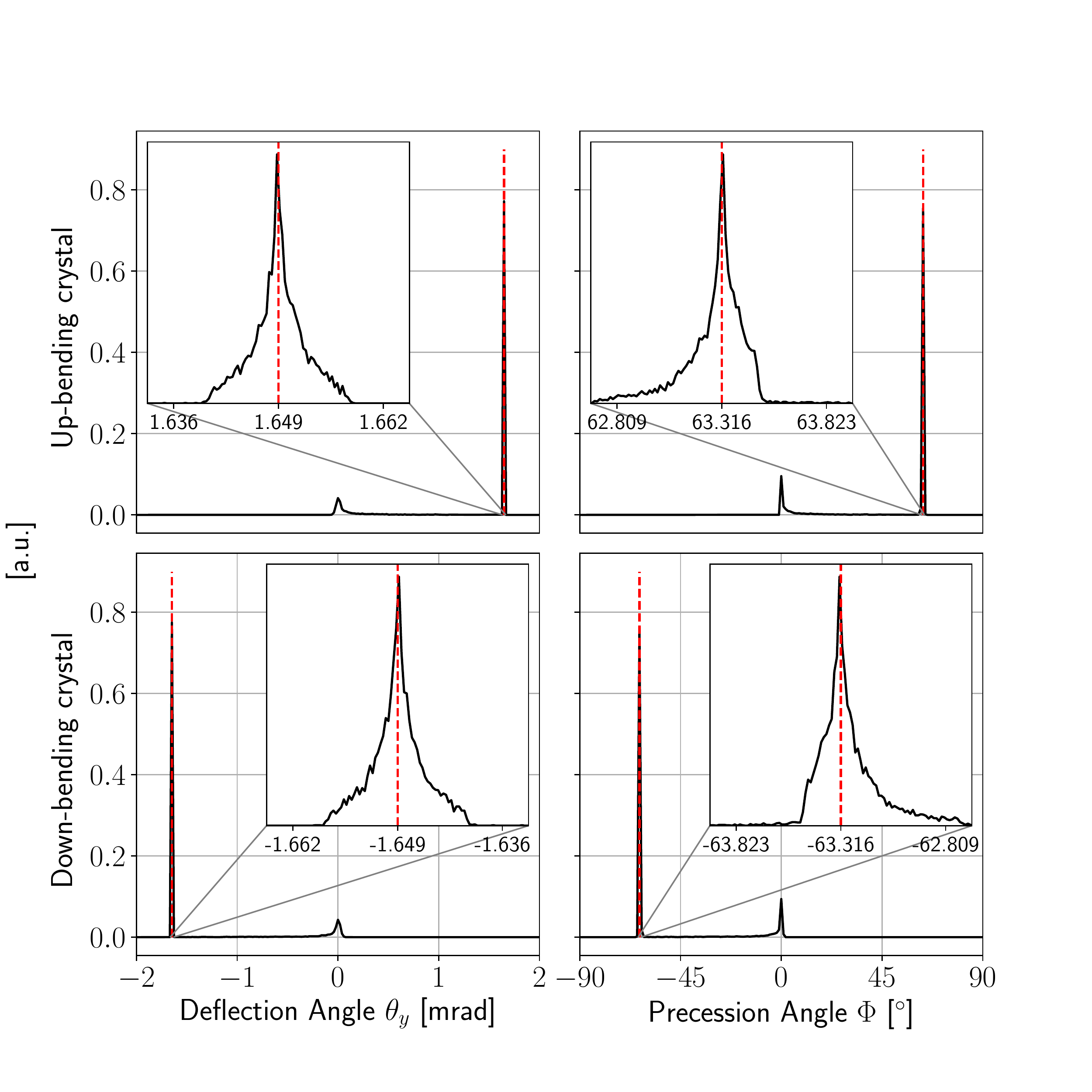}
\caption{\label{fig:E761} 
  Distributions of the trajectory deflection angle and spin precession angle
  for \Sigmap baryons of $375$\gevc momentum interacting with $4.5$\cm long (top) up-bent and (bottom) down-bent crystals,
uniformly bent along the $(111)$ plane at $\pm 1.649$\mrad angle.
Similar crystals were used for the E761 experiment at FNAL~\cite{Chen:1992wx}.}
\end{figure}

The \geant application for spin precession has been validated against the solely available experimental
data provided by the E761 experiment at FNAL~\cite{Chen:1992wx}. In that experiment, two $4.5$\cm long \Si crystals bent along the $(111)$ plane
were exposed to a \Sigmap beam with $375$ GeV/c momentum. The deflection angle of the two crystals were $+1.649\mrad$
and $-1.649\mrad$, with measured precession angles of $-72^{\circ} \pm 26^{\circ}$ and $+51^{\circ} \pm 23^{\circ}$, respectively.
As expected, the spin in the two crystals precesses in opposite directions.
The average of experimental values $60^{\circ} \pm 17^{\circ}$ is consistent with the predicted value of $62^{\circ} \pm 2^{\circ}$.   
%
%
A uniformly bent crystal with the E761 geometrical parameters has been implemented within \geant and exposed to a monochromatic and 
perfectly collimated \Sigmap beam with $375$\gevc momentum.
Figure~\ref{fig:E761} shows the distributions of the trajectory deflection angle and of the spin precession angle for both the 
up- and down-bending crystals. A precession of $+63.3^{\circ} \pm 0.2^{\circ}$ and $-63.3^{\circ} \pm 0.2^{\circ}$ was obtained for the two cases, respectively,
in good agreement with the predicted values of $\pm 63.0^\circ$.

\begin{figure}
\includegraphics[width=1\columnwidth]{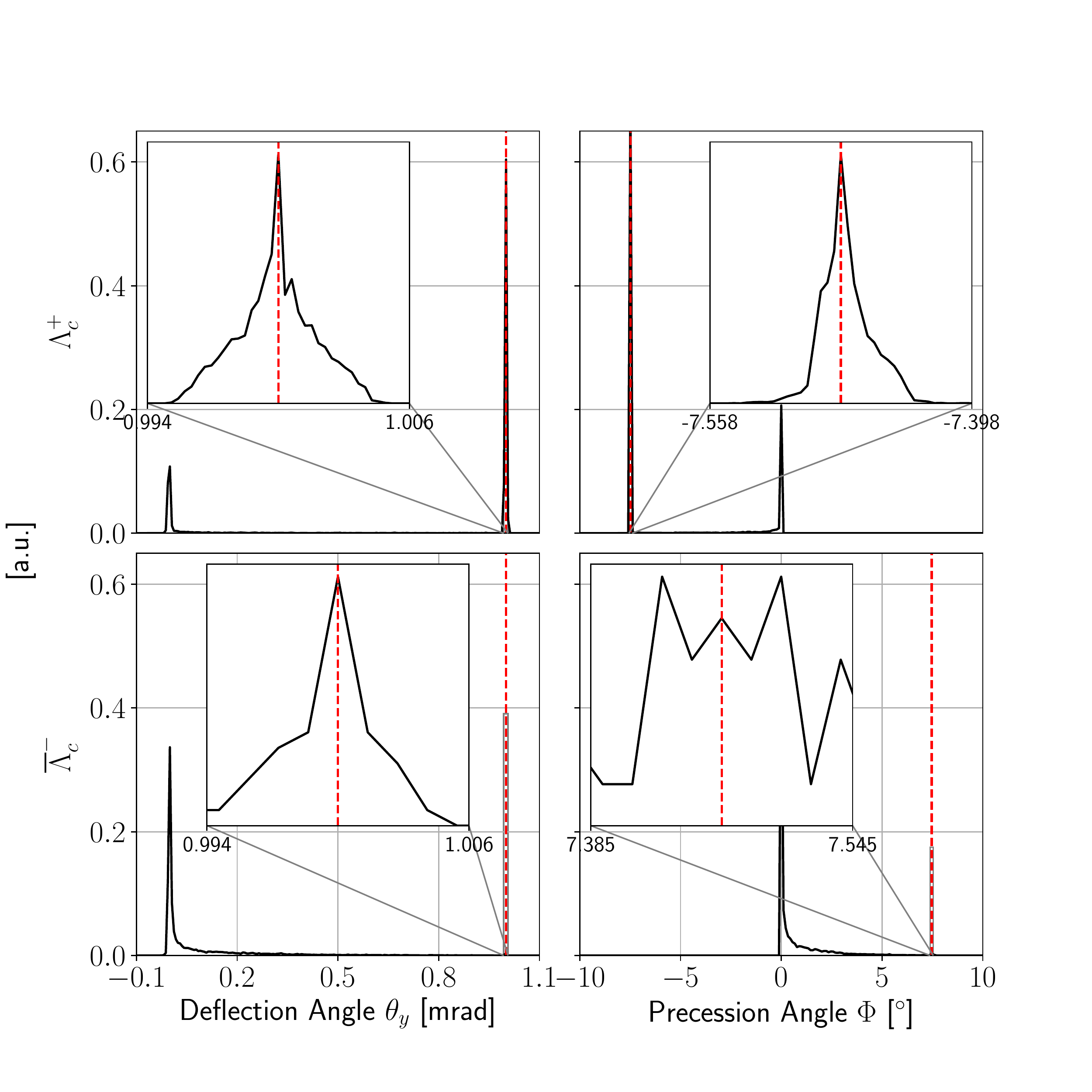}
\caption{\label{fig:lambdac} 
  Distributions of the trajectory deflection angle and spin precession angle for
  (top) \Lc  and (bottom) \Lcbar  baryons of $1$\tevc momentum
interacting with a $1$\cm long crystal uniformly bent along the $(110)$ plane at $1$\mrad angle.}
\end{figure}

%
%
%
%
%
The  same \geant toolkit can also be used to simulate the spin precession of other positive and negative particles in a bent crystal, to compare
with the expected analytical values. For this purpose, a $1$\cm long \Si crystal bent along the $(110)$ plane by a $1$\mrad bending angle
has been used. A beam of short-lived particles with no angular divergence is generated in the simulation immediately before the
crystal and the precession angle at the end of the crystal is evaluated. 
Table~\ref{tab:precession_g_d} shows the simulation results for pairs of short-lived particles and their antiparticles, \ie~$\Lc/\Lcbar$, 
$\Xim/\Xip$, $\Omegam/\Omegap$, and $\Xibm/\Xibp$, in presence of MDM and EDM, respectively.
Figure~\ref{fig:lambdac} shows the distribution of the measured trajectory deflection angles and spin precession angles
for the \Lc/\Lcbar case.
\begin{table*}[ht]
\centering
\caption{
Average spin precession angle ($\Phi_{\rm sim}$) and EDM polarization component ($s_{x,\rm sim}$) obtained from \geant simulation compared to the expected
values ($\Phi_{\rm exp}$ and $s_{x,\rm exp}$, respectively), due to the gyromagnetic factor of the particle expressed as $g'=(g-2)/2$ and the 
gyroelectric factor $d = 5\times10^{-2}$, along with the mean channeling deflection efficiency ($\effc$),
for different 1\tevc particles impinging on a $1$\cm long \Si crystal bent along the $(110)$ plane at
$1$\mrad angle.
The normalization of the polarization vector $\mathbf{s_0}$ has been taken unity, \ie $s_0=1$.
\label{tab:precession_g_d}}
\makebox[\textwidth]
{\begin{tabular}{lcccccccc}
\toprule
    Particle &   $g'$ & $\Phi_{\rm exp}$ [$^{\circ}$] & $\Phi_{\rm sim}$ [$^{\circ}$] & $s_{x,\rm exp}$ & $s_{x,\rm sim}$ & $\effc$ [$\%$] \\
\midrule
    \Lc     & $-0.30$    & $-7.518$    & $\phd-7.474\pm0.015$  & $\phm7.17\times10^{-4}$ & $(\phm7.19\pm 0.03)\times10^{-4}$ & $71.0\pm0.08$  \\
    \Lcbar  & $\phm0.30$ & $\phm7.518$ & $\phz\phm7.59\pm0.07$ & $-7.17\times10^{-4}$  & $(-7.20\pm 0.13)\times10^{-4}$      & $0.51\pm0.07$ \\
    \Xim    & $-1.92$    & $-83.09$    & $-83.0\pm0.9$       & $\phm1.132\times10^{-2}$ & $\phm(1.145\pm 0.020)\times10^{-2}$   & $0.47\pm0.07$ \\
    \Xip    & $\phm1.92$ & $\phm83.09$ & $\phm83.21\pm0.23$  & $-1.132\times10^{-2}$    & $(-1.149\pm 0.005)\times10^{-2}$      & $70.6\pm0.08$ \\
    \Omegam & $-2.20$    & $-75.38$    & $-75.4\pm0.6$       & $\phm8.50\times10^{-3}$  & $\phm(8.51\pm 0.12)\times10^{-3}$  & $0.39\pm0.6\phz$ \\
    \Omegap & $\phm2.20$ & $\phm75.38$ & $\phm75.53\pm0.15$  & $-8.50\times10^{-3}$     & $(-8.51\pm 0.03)\times10^{-3}$     & $70.9\pm0.8\phz$ \\
    \Xibm   & $-1.38$    & $-13.65$    & $-13.64\pm 0.14$      & $\phm5.154\times10^{-4}$ & $\phm(5.15\pm 0.10)\times10^{-4}$ & $0.51 \pm 0.07$ \\
    \Xibp   & $\phm1.38$ & $\phm13.65$ & $\phm13.78\pm 0.03$   & $-5.154\times10^{-4}$    & $(-5.167\pm 0.021)\times10^{-4}$  & $71.0 \pm 0.8\phz$ \\
\bottomrule
\end{tabular}}
\end{table*}

%
\section{The experiment}
\label{sec:setup}

%
%
%

The MDM and EDM information can be extracted using Eqs.~(\ref{eq:Phi}),~(\ref{eq:EDM_precession}) and~(\ref{eq:MDM_precession}), 
from the  measurement of the spin polarization of channeled baryons at the exit of the crystal,
via the study of the angular distribution of final state particles.
For \Lc decaying to two-body final states such as $f = \Deltares^{++}\Km$, $p\Kstarz$, $\Lz(1520)\pip$ and $\Lz\pip$, the angular distribution is described by
\begin{equation}
\frac{dN}{d\Omega}\propto 1+\alpha_f {\mathbf s}\cdot \hat{\mathbf k}, 
\label{eq:Lc_AngDist}
\end{equation}
where $\alpha_f$ is a parity-violating coefficient depending on the final state $f$,
$\hat{\mathbf k}$ the 
direction of the final state baryon
in the \Lc helicity frame,
$\Omega$ the corresponding solid angle,
and ${\bf s}$ the \Lc polarization vector. 
Equation~(\ref{eq:Lc_AngDist}) holds for all spin-1/2 decays into two-body final states with spins
$1/2+0$, $1/2+1$ and $3/2+0$.

\begin{figure*}[htb]
\centering
{\includegraphics[width=0.9\linewidth]{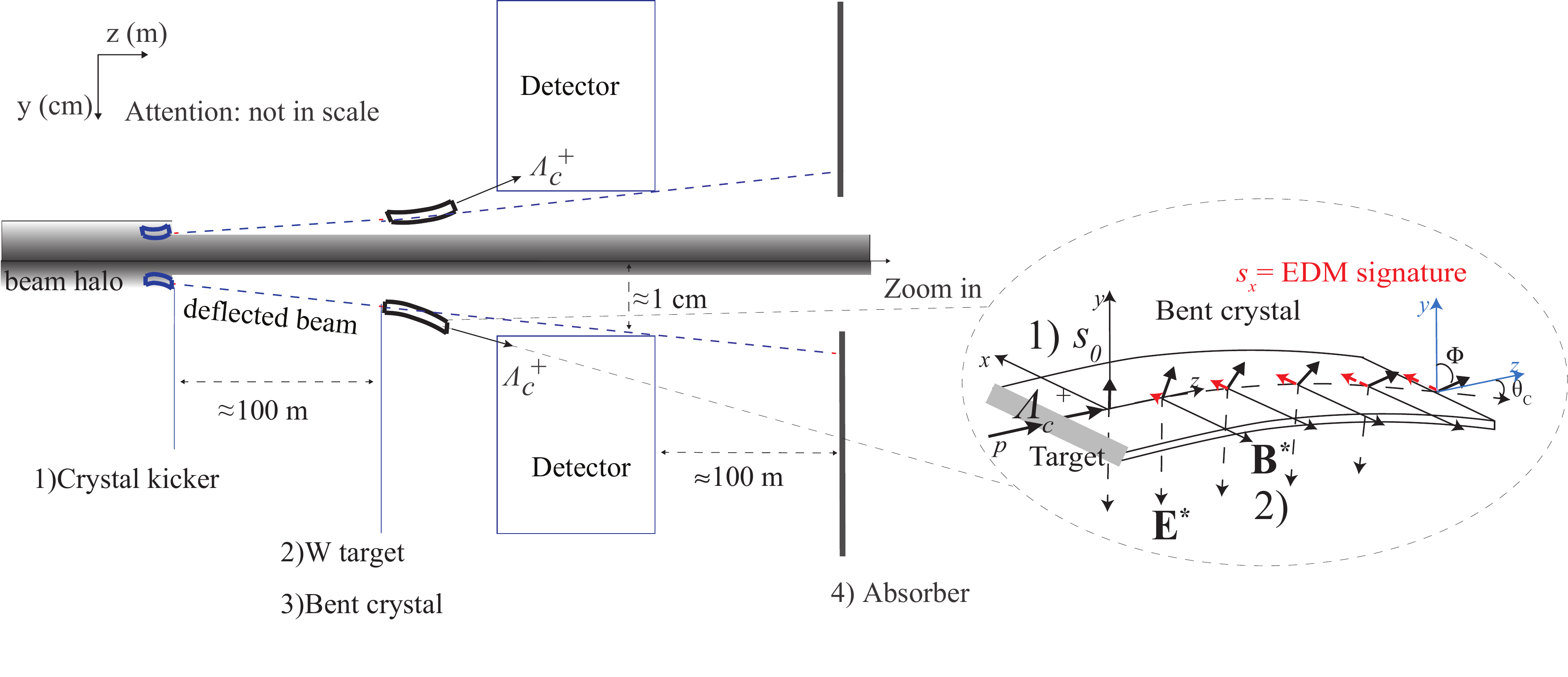} }
\caption{
Conceptual layout of the fixed-target setup shown in side view with down- and up-bending crystals.
  The zoom in shows the spin precession in the down-bending crystal for channeled \Lc baryons.}
\label{fig:FT_Layout}
\end{figure*}

The initial polarization $s_0$ would require in principle the measurement of the angular distribution 
for unchanneled baryons. In practice, however, this is not required since the measurement of the 
three components of the final polarization vector for channeled baryons allows a simultaneous determination of 
$g$, $d$ and $s_0$, up to discrete ambiguities, as discussed in Appendix~\ref{app:ambiguities}. These, in turn, 
can be resolved exploiting the dependence of the angular distribution with the \Lc boost $\gamma$.

\subsection{Possible experimental layout}
\label{sec:layout}

The possible experimental layout is based on the double crystal scheme~\cite{Burmistrov:2194564} sketched in Fig.~\ref{fig:FT_Layout}. It consists of four main elements:
\begin{enumerate}

\item
two crystal kickers positioned about $100\m$ upstream of the target for up and down deflection at angle
$\approx 100\murad$ of the 7\tev protons from the \lhc beam halo. This technique has been demonstrated
to be feasible without affecting the \lhc beam lifetime~\cite{Scandale:2016krl};
%
\item
 two amorphous \W targets about $0.5\cm$ thick intercepting the deflected proton beam where charm, beauty and 
strange baryons are produced.
The fixed target has to be installed in front of the detector, as close as possible to obtain good vertex resolution;
\item
up- and down-bending crystals to induce opposite spin precession to channeled baryons. 
The use of two crystals with opposite bendings is crucial to prove the robustness of the results and control systematic uncertainties.
The \W target should be attached to the crystal to maximize the yield of channeled baryons;
\item
two absorbers 
positioned downstream of the detector to stop the deflected proton beam
and background particles induced by the interactions with the target and crystal materials~\cite{Redaelli:PBC}.
%
%
\end{enumerate}

To protect against radiation damage and minimise detector occupancies, the design has to guarantee that
non-interacting protons and unchanneled particles
follow the beam pipe towards the absorbers.

Despite its challenges, the setup is based on two key elements already existing and tested successfully 
at the \lhc: high-purity bent crystals and high-accuracy positioning systems (goniometers).
Two bent crystal types with different characteristics are required.
The crystal kicker is very similar to that one tested at the \lhc~\cite{Scandale:2016krl}, of 4 \mm length
and $\approx100\murad$ bending angle;  the second crystal should have larger angle,
order $10 \mrad$, as discussed later.
The remotely controlled goniometers equipped with the bent crystals are mounted on standard collimation supports 
and make use of fast plug-in technology, which ensures fast handling of the object.
They are based on a piezoelectric actuator and feature angular resolution
of $0.1\murad$ and linear resolution of $5\mum$. This is necessary to align the crystal kicker with respect to the beam halo,
and position the long-bending crystal to intercept the deflected beam.  

Two possible configurations have been considered for the fixed-target and detector setup, referred to hereafter as \Sa and \Sb.
The former is based on the upgraded \lhcb detector~\cite{LHCb-TDR-012}, which will become operational in 
2021 and will run for the rest of 
the decade (\lhc Run 3 and Run 4), whereas the latter is an hypothetical dedicated detector considered to function
at even higher luminosities and providing an angular coverage to minimise the crystal bending angle.

\lhcb is a single-arm forward spectrometer~\cite{Alves:2008zz,LHCb-DP-2014-002} 
dedicated to the study of particles containing \bquark or \cquark quarks at \lhc. The detector tracking
system consists of three main devices: a vertex locator (VELO) surrounding the \pr\pr interaction region and a large-area 
detector (TT) upstream of the dipole magnet inducing an integrated field of about 4 T\m, and three stations (T1-T3) downstream 
of the magnet. Particle identification is provided by two ring-imaging Cherenkov detectors, a calorimeter system and muon chambers. 
With the upgrade, 
most of the sub-detectors will be replaced~\cite{LHCb-TDR-013,LHCb-TDR-015,LHCb-TDR-014} 
and a full software based trigger will become operational~\cite{LHCb-TDR-016}, providing significantly increased efficiencies in hadronic final 
states and allowing the experiment to operate at higher luminosities.
The fixed-target like geometry combined with the capability to reconstruct with good efficiency
highly-boosted baryons makes of \lhcb the most suitable detector for this proposal.

\begin{figure}[htb]
\centering
\includegraphics[width=0.27\textwidth]{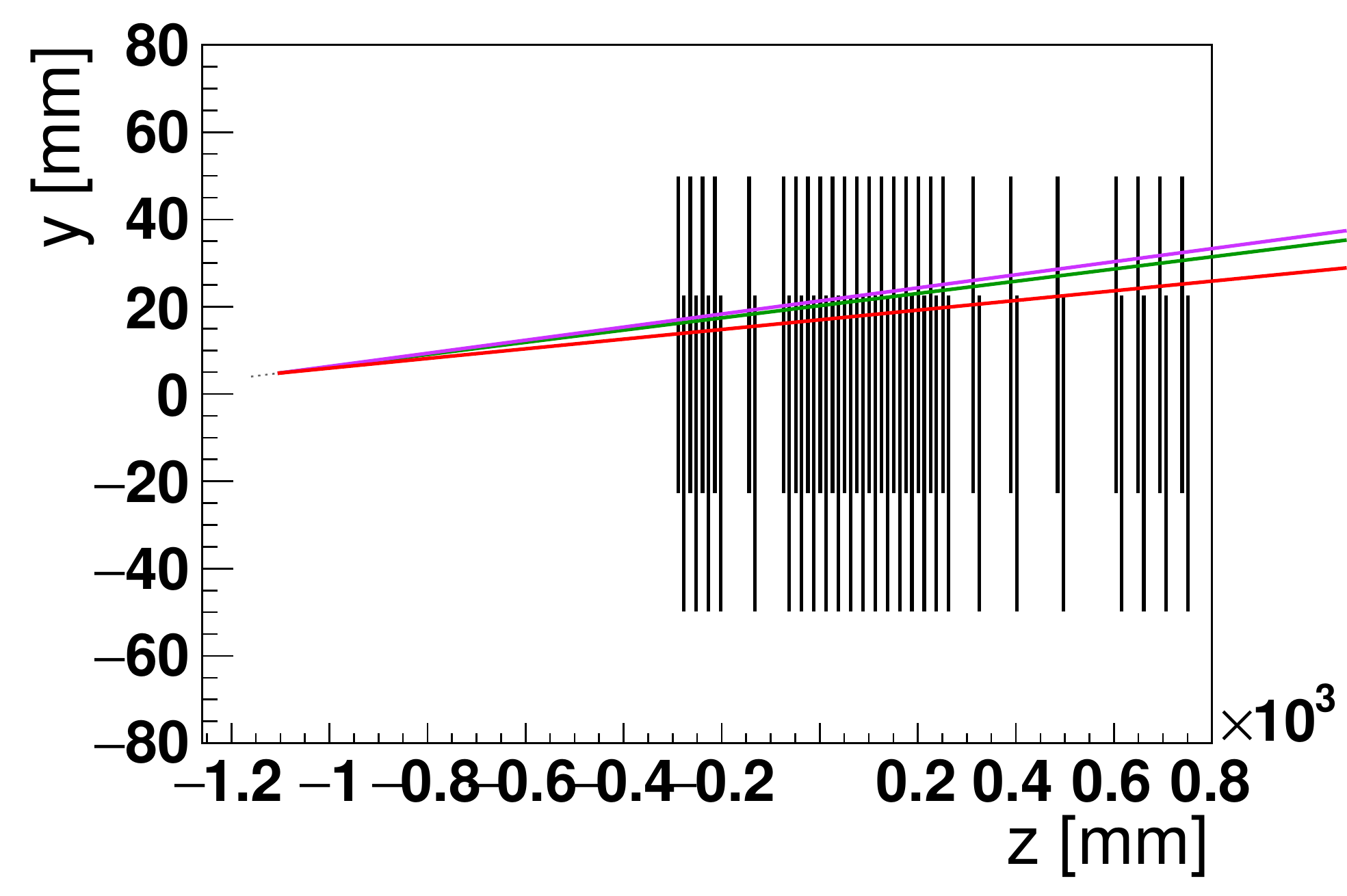}
\includegraphics[width=0.19\textwidth]{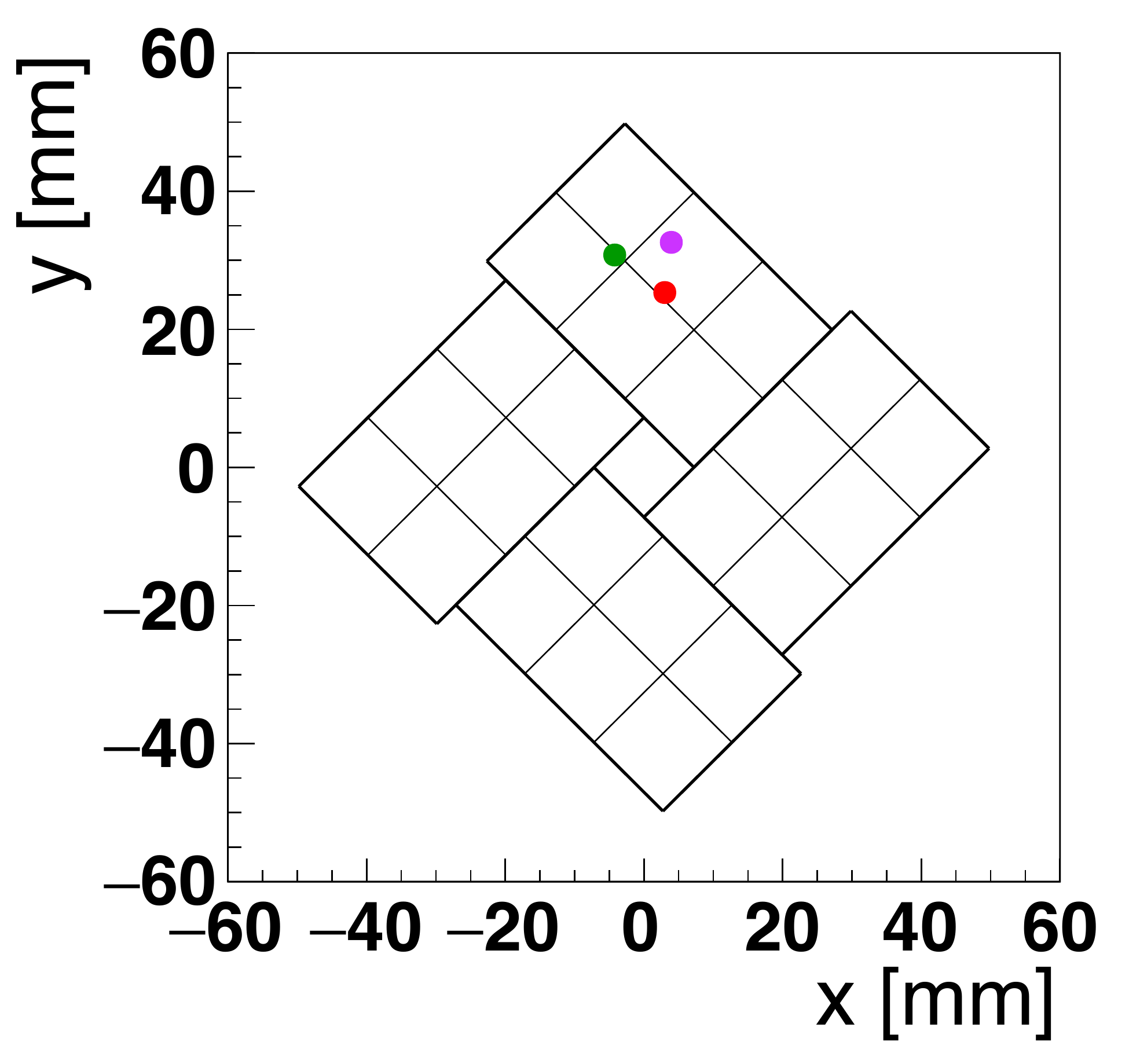} \\
\includegraphics[width=0.27\textwidth]{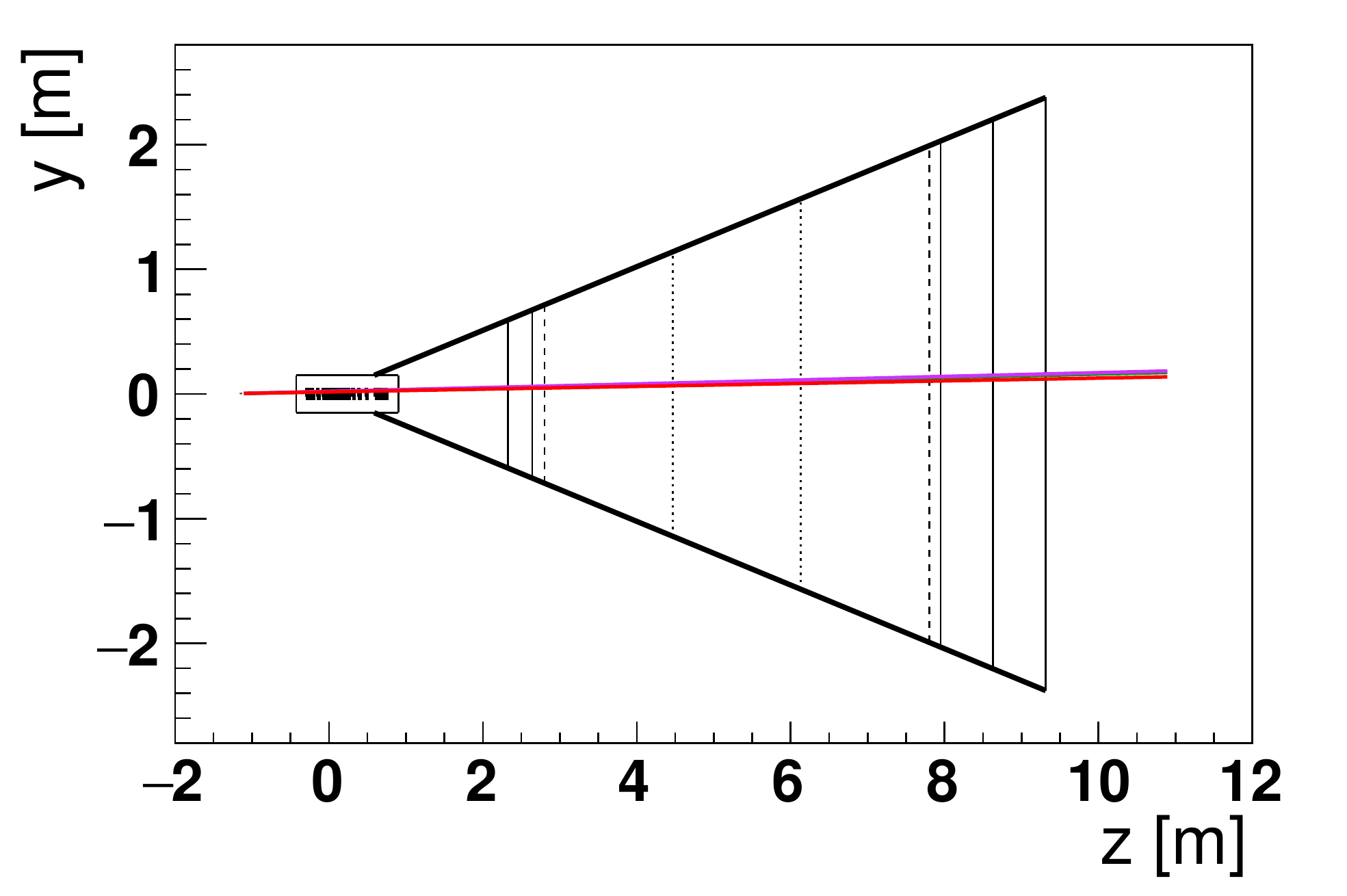}
\includegraphics[width=0.19\textwidth]{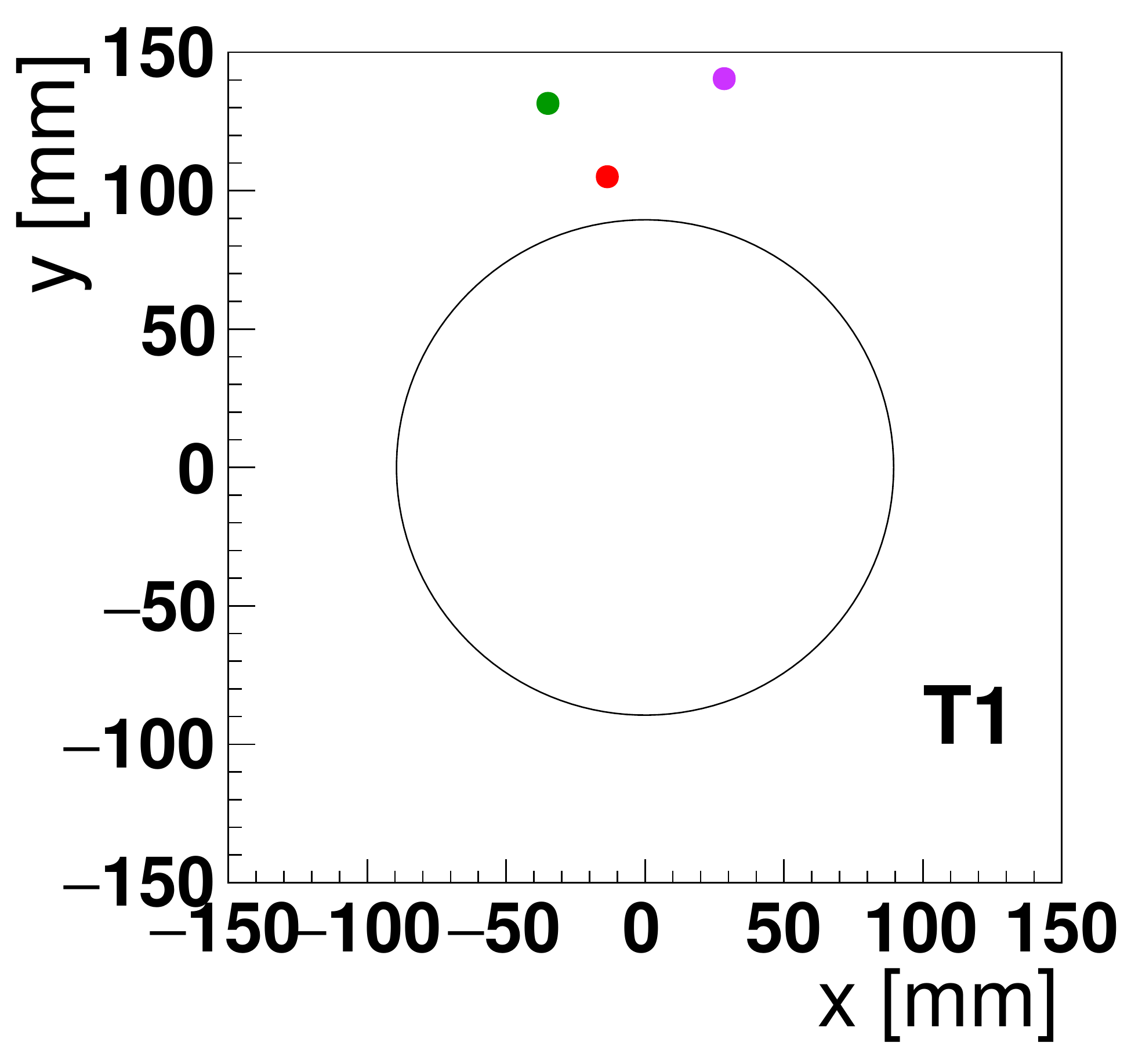}
\caption{\label{fig:EventDisplay} 
Sketch of a fixed-target $\Lc\to\pr\Km\pip$ event generated with \pythia at 
position $(0,0.4,-116)\cm$ with a crystal bending $\theta_C=14\mrad$, as seen by the 
simplified \lhcb detector geometry model based on Refs.~\cite{LHCb-TDR-013,LHCb-TDR-015}
with a conservative beam clearance of the downstream stations of 30\mm. 
The first VELO module is located at $z \approx -29$\cm, 
 upstream of the nominal \lhc collision point at $z=0$.
The points represent the hits of the proton (green), pion (violet) and kaon (red) tracks overlaid in the (top left) side view of the VELO, 
(top right) front of the last VELO module, (bottom left) schematic side view of the whole \lhcb detector, 
and (bottom right) central area of the T1 station.
Events are considered within acceptance when they cross at least three VELO modules and the three T stations.
The track bending due to the \lhcb dipole magnet is taken into account.
}
\end{figure}

To minimise the impact on the interaction region, in the \Sa scenario the fixed-target setup is positioned 
outside of the VELO vessel container~\cite{LHCb-TDR-013}, $\approx 1.16$\m upstream of the nominal \pr\pr collision point and 
$\approx 0.87$\m upstream of the first VELO module.
At this position, a minimal crystal bending angle of about 
14\mrad 
is required for channeled
baryons to be deflected inside the detector fiducial volume. 
This follows from a 
Monte Carlo simulation of fixed-target events using \pythia~\cite{Sjostrand:2006za} together with a simplified geometrical model 
of the upgraded \lhcb detector, as sketched in Fig.~\ref{fig:EventDisplay}. 
%
For the \Sa scenario, a beam intensity of $5\times10^8\proton/\sec$ impinging the target,
and an overall data taking efficiency of 50\% is assumed.
In this case in six weeks of dedicated detector operations,
spanned over several years during the next decade, it would be possible to achieve a statistics of 
about $10^{15}$ protons on target (\pot). Accurate studies of the attainable proton flux
are in progress~\cite{Redaelli:PBC}, which has to be compliant with the \lhc beam lifetime,
machine operations and protection.
Detector occupancies at such rates are expected to be manageable, as discussed later. 

\begin{figure}[htb]
\centering
\includegraphics[width=0.45\textwidth]{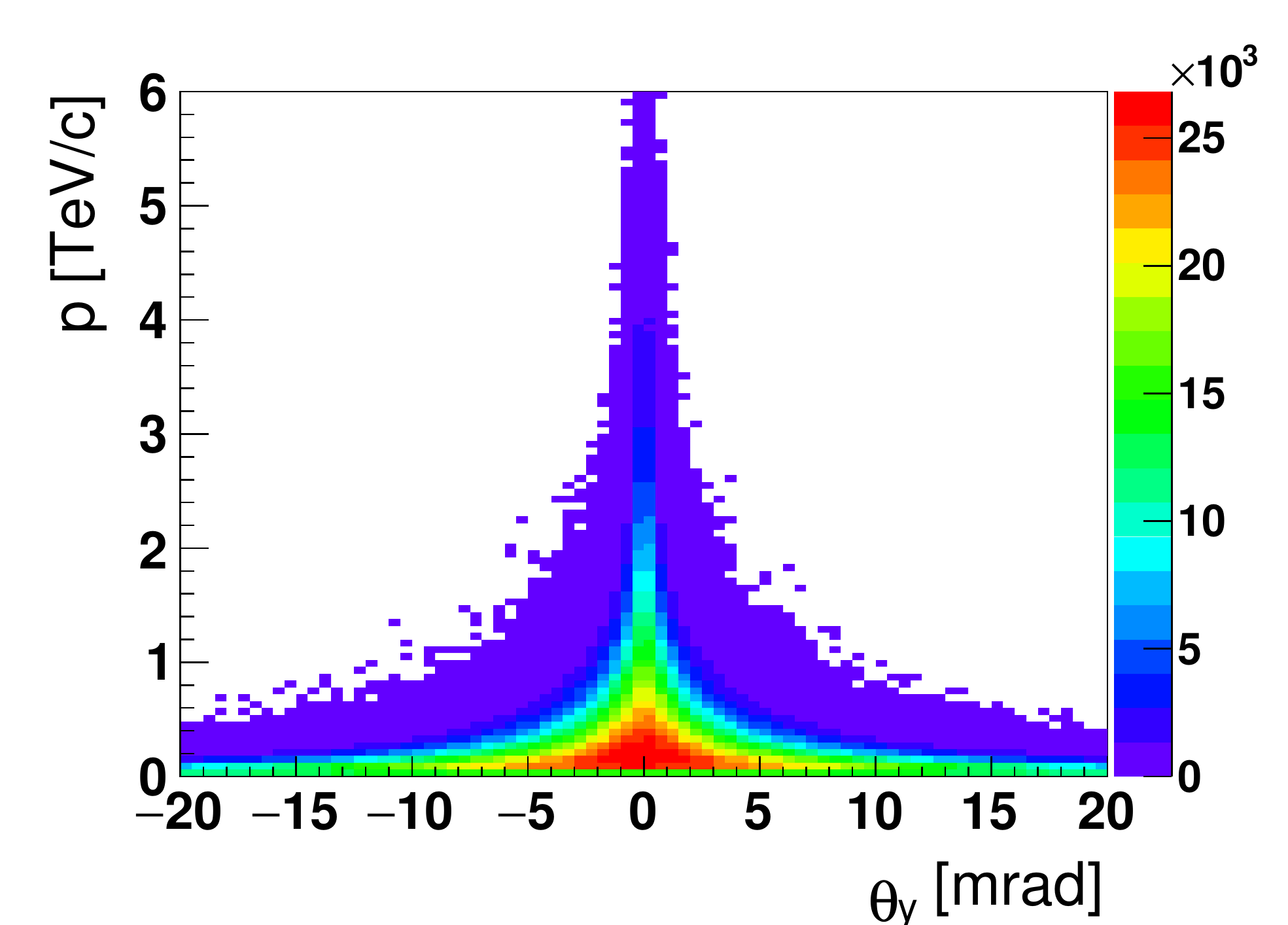}
\caption{\label{fig:ThetavsP_all} Distribution of polar angle $\theta_y$ versus momentum for \Lc baryons produced in 7\tev proton beam collisions on protons at rest using \pythia. 
}
\end{figure}

%
An increase in proton fluxes, \eg at High Luminosity \lhc (HL-\lhc)~\cite{Apollinari:2017cqg},
combined with the design of a dedicated detector capable to afford the higher occupancy levels 
and longer data taking periods, could potentially offer the opportunity for \Sb scenario
to integrate $\sim 10^{17}$ \pot.
Such detector could also extend the angular coverage at larger pseudorapidity to minimise the crystal bending angle, 
increasing the channeling efficiency,
and operate closer to the fixed-target to provide enhanced vertex resolution. 
Nevertheless, a minimal bending angle of about 5\mrad is needed to guarantee a good separation between channeled and unchanneled particles.
This value can be inferred from the angular distribution of 
\Lc baryons produced by 7\tev protons impinging on the fixed-target,
which are highly collimated along the incident proton beam
direction, and are isotropically distributed over the azimuthal angle. The polar angle 
that defines the emission cone is $\propto \gamma^{-1} \approx 1 \mrad$. 
The corresponding \pythia distribution of polar angle versus momentum, illustrated in Fig.~\ref{fig:ThetavsP_all}, 
shows that for $|\theta_y|>5\mrad$ and momentum higher than $\approx 1$\tev there are practically no unchanneled \Lc baryons.

%
%
%
\subsection{Crystal parameters}
\label{sec:crystalparams}
In Sec.~\ref{sec:channeling} is discussed how the channeling efficiency depends on the crystal
parameters and on the momentum range of the particles.
In the following, the length $L$ and bending angle $\theta_C$ for \Si and \Ge crystals are estimated
maximizing the sensitivity of the experiment to the electromagnetic moments,
while taking into account detector acceptance.
The optimization has been performed using fixed-target $\Lc\to\pr\Km\pip$ events, produced in
7\tev proton beam collisions on protons at rest using \pythia. 
The \Lc channeling has been simulated using a parameterisation 
based on current theoretical description and channeling measurements, following Ref.~\cite{Biryukov1997}.

A particle entering the crystal is channeled when its polar angle $\theta_y$ lies within the
$(-\theta_L,\theta_L)$ interval, 
where $\theta_L$ is the Lindhard angle, introduced in Sec.~\ref{sec:channeling}.
%
For $p\approx 1$\tev in Si (Ge) this angle is about 6 (7)\murad, about three orders
of magnitude smaller than the angular divergence of \Lc baryons shown in Fig.~\ref{fig:ThetavsP_all}. Therefore, the trapping efficiency 
\efft
is largely
dominated 
by the angular opening of the baryons produced in the target.
This imposes the crystal to be directly attached to the \W target.
The per-event 
deflection
efficiency is parameterised as
\begin{equation}
\label{eq:channeling_eff}
\effc = (1-\eta_{c})^2 e^{-\theta_C / [ \theta_d \eta_{c} (1-\eta_{c})^2 ]},
\end{equation}
where dechanneling losses inside the bent crystal are described by the 
factor
$(1-\eta_{c})$ introduced in Sec.~\ref{sec:channeling}, 
which accounts for the shortening of the dechanneling length with respect
to a straight crystal, 
and 
$\theta_d$, which is the ratio of $L_d$ in a straight crystal to $R_c$.
%
The per-event critical radius, 
$R_{c} = p c/U'(x_c)$, 
must be below the crystal curvature radius $R=L/\theta_C$,
where $U'(x_c)$ is the interplanar electric field at the critical transverse coordinate, below which
the particle is lost from channeling mode.
For Si 110 (Ge 110), $x_c \approx 0.885~(0.915)$~\AA\ and $U'(x_c) \approx 5.7~(10)$\gev/\cm~\cite{Biryukov1997}.
 
\begin{figure}[htb]
\centering
\begin{tabular}{ c c }
\includegraphics[width=0.23\textwidth]{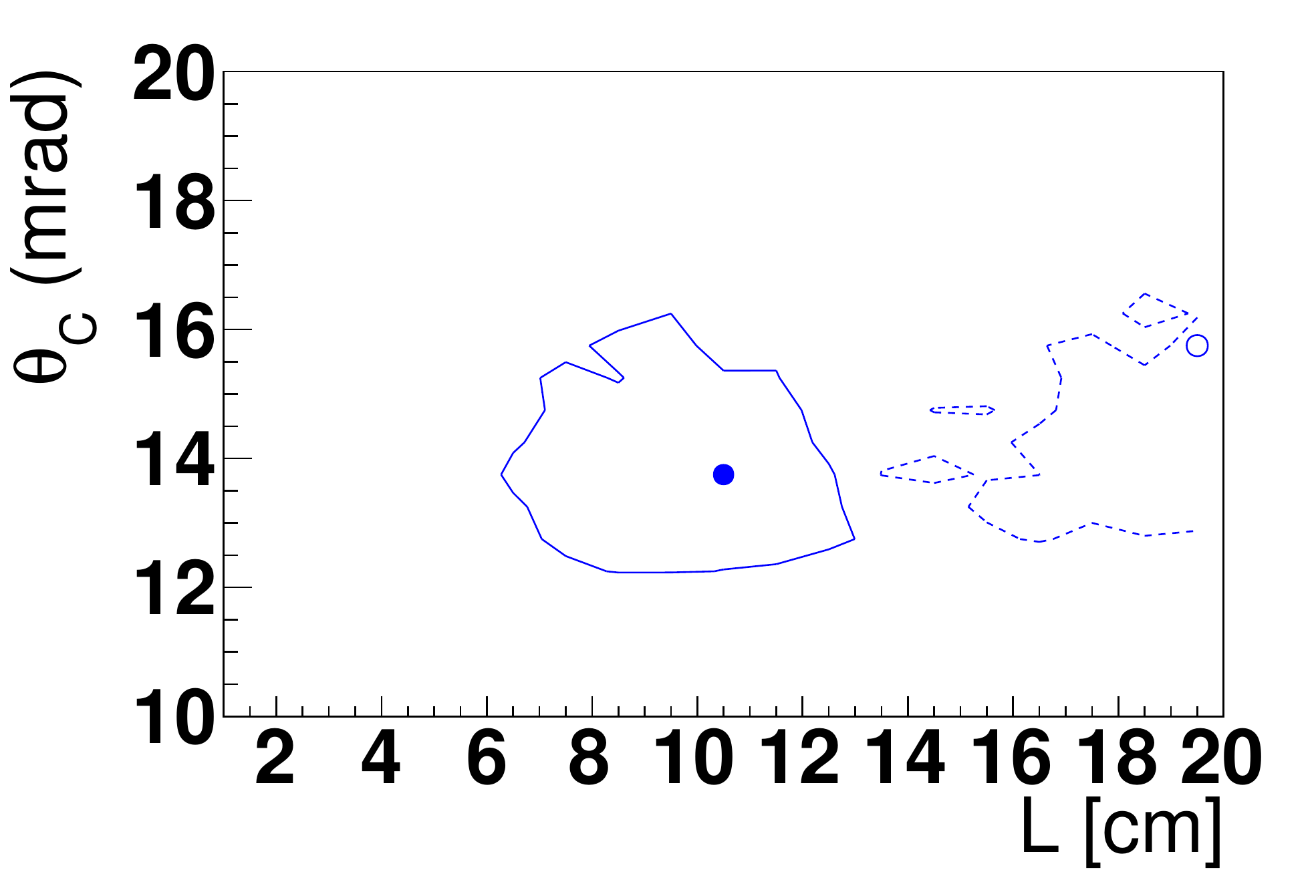} &
\includegraphics[width=0.23\textwidth]{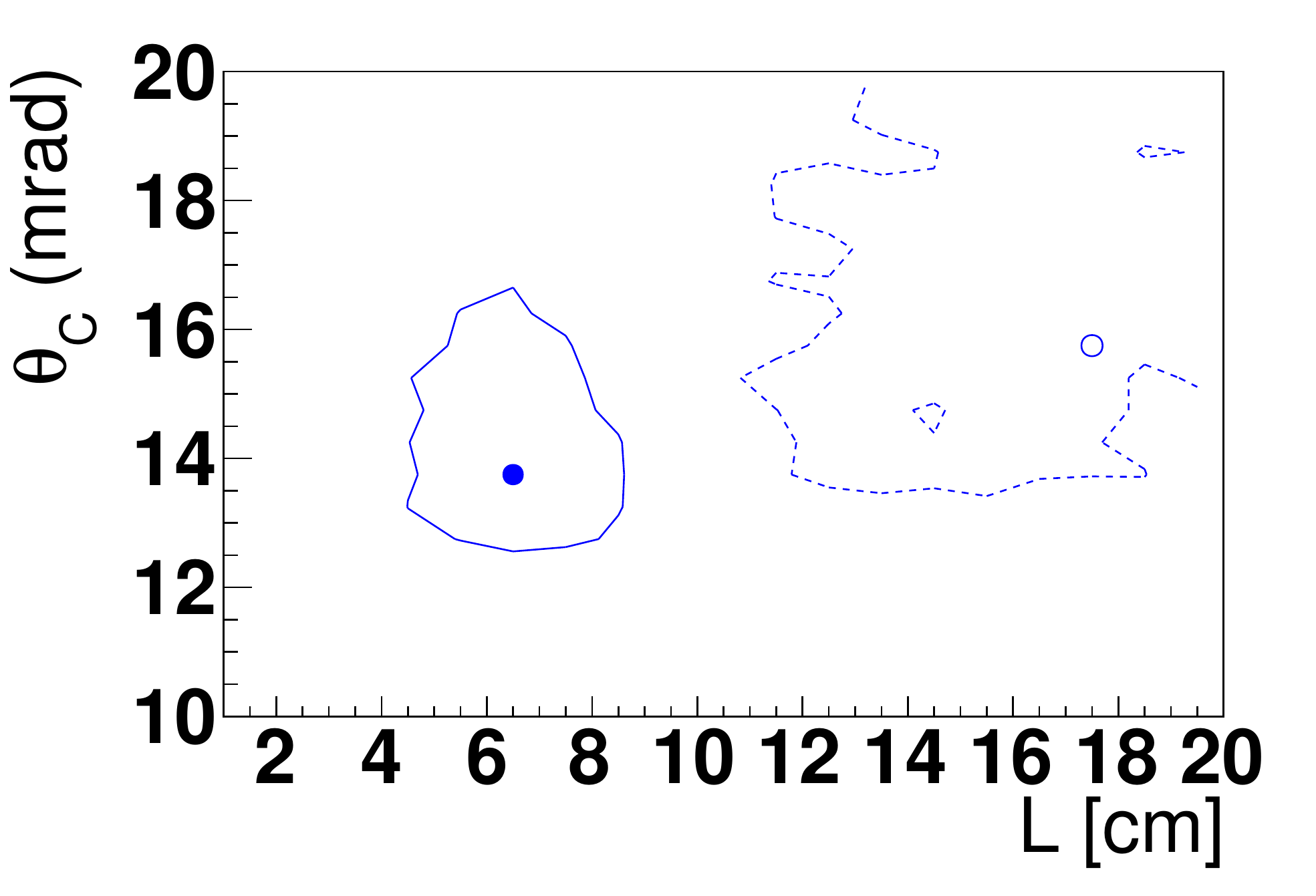} \\
\includegraphics[width=0.23\textwidth]{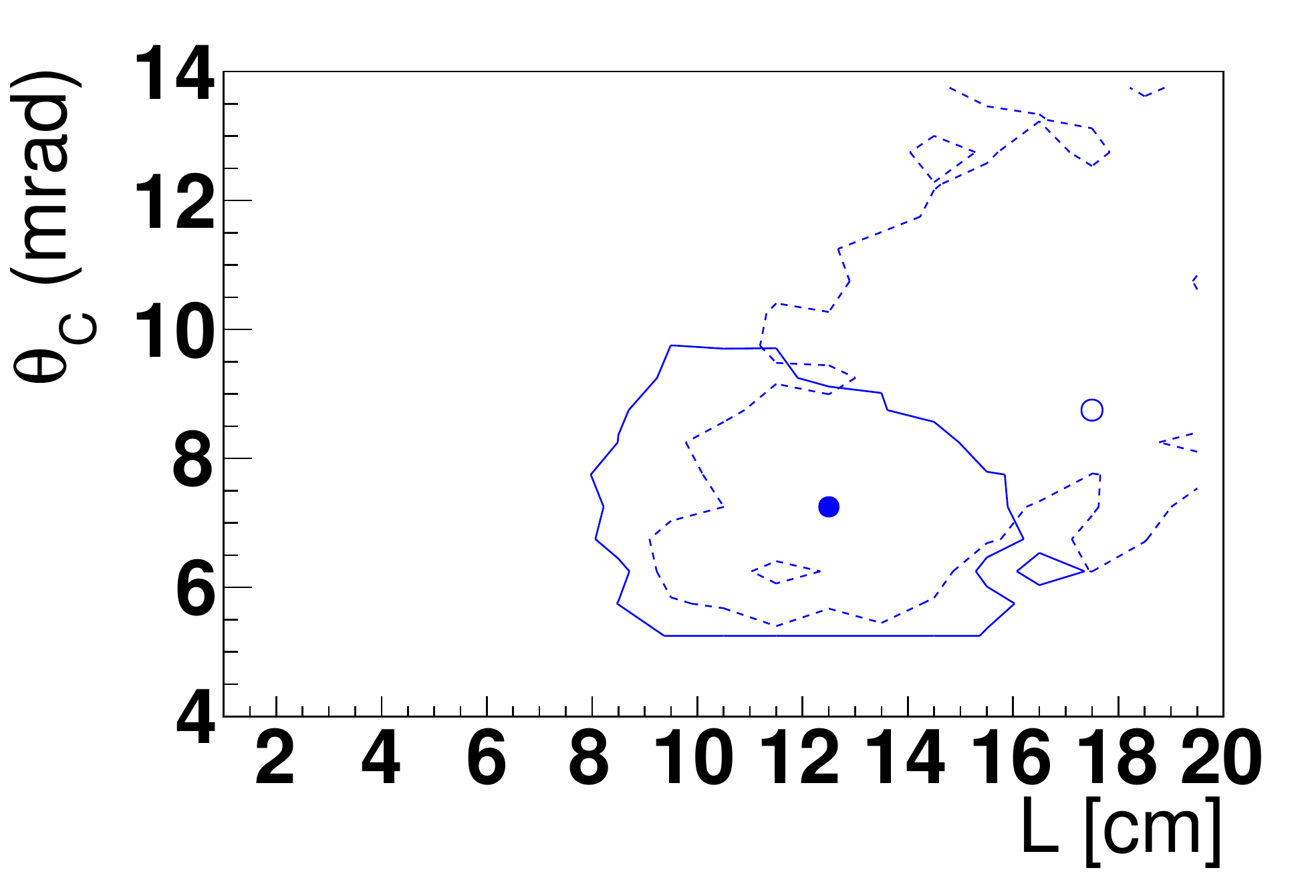} &
\includegraphics[width=0.23\textwidth]{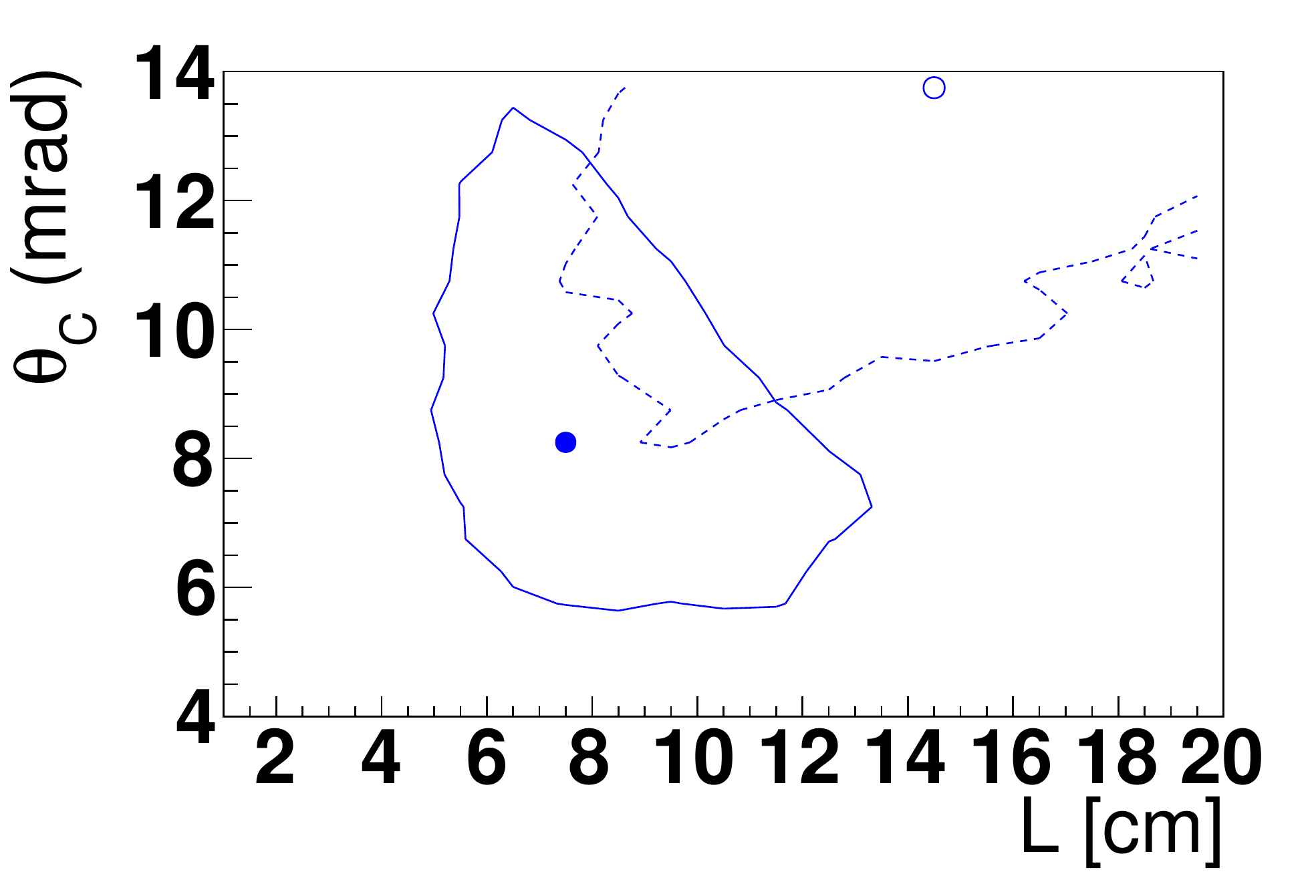} \\
\end{tabular}
\caption{\label{fig:d_g_error}
Regions of minimal uncertainty of $d$- and $g$-factors as a function of the crystal parameters $L$ and $\theta_C$, for (left) Si and (right) Ge
in (top) \Sa and (bottom) \Sb scenarios.
The markers and continuous (dotted) lines represent the minimum uncertainty and regions whose uncertainties on $d$ ($g$) are increased by 20\% 
with respect to the minimum, respectively.
}
\end{figure}

A scan in the $(L,\theta_C)$ plane is performed to determine the minimal error on the $d$ and $g$ factors separately.
Since there is a wide momentum distribution for the channeled \Lc baryons, we generate and fit
pseudo-experiments using a conditional
probability density function constructed with the angular distribution in Eq.~\eqref{eq:Lc_AngDist}.
Following the discussion at the end of Sec.~\ref{sec:layout}, we require the momentum of the \Lc baryons to be higher than 800~\gevc.
The dependence of the spin polarization is obtained from Eqs.~\eqref{eq:EDM_precession} and~\eqref{eq:MDM_precession},
assuming $d=0$, $g'=(g-2)/2=-0.3$, $\alpha_f = -0.67$ and $s_0 = -0.6$, as described in more details in Sec.~\ref{sec:sensitivity}.
Figure~\ref{fig:d_g_error} shows regions whose uncertainties on $d$ ($g$) are increased by 20\% with respect to the minimum,
for Si and Ge in both \Sa and \Sb scenarios.
The gyromagnetic ratio prefers higher $L$ values, as $\sigma_g \propto 1/\gamma$. 
These wide regions provide the optimal parameters summarized in Table~\ref{tab:crysparam}, chosen around the minimum 
of the gyroelectric factor.

\begin{table}[htb]
\caption{
Bent crystal parameters for \Si and \Ge optimized using charm baryon decays, 
for the two possible experimental scenarios under consideration. 
The intervals give approximate regions whose uncertainties on the $d$ factor are increased by 20\% with respect to the minimum,
whereas the central values are chosen for the sensitivity studies discussed in Sec.~\ref{sec:sensitivity}.
\label{tab:crysparam}}
\begin{center}
\begin{tabular}{lcccc}
\toprule
                   & \multicolumn{2}{c}{S1} & \multicolumn{2}{c}{S2} \\
\midrule
                   & \Si     & \Ge          & \Si     & \Ge \\
\cmidrule{2-5}
$L$ $[\cm]$          &  $[6,12]$  & $[4,8]$  & $[8,15]$ &  $[5,12]$  \\
                     &  $7$       & $5$      & $12$     &  $7$  \\
$\theta_C$ $[\mrad]$ &  $[13,16]$  & $[13,16]$  & $[5,9]$ &  $[6,12]$ \\
                     &  $14$       & $15$       & $7$     &  $8$ \\
%
$R/R_c$              &  $2.85$        & $3.33$      & $9.77$    &  $8.75$ \\
\bottomrule
\end{tabular}
\end{center}
\end{table}

The parametric approach adopted above to account for dechanneling losses has been validated with 
\geant simulations, discussed in Secs.~\ref{sec:channeling} and~\ref{sec:spin_geant4}, 
using 1\tev \Lc baryons with no angular divergence.
For a 7 (5)\cm long \Si (\Ge) crystal bent along the (110) plane by a 14 (15)\mrad bending angle, 
the channeling deflection efficiency  $\eps_c$ is found to be $37.8\pm 0.6$ ($31.7\pm 0.6$)\%.
In this case the ratio of the optimal crystal bending radius to the critical bending
radius, $R/R_c$, is $2.85$ ($3.33$) for the \Si (\Ge) crystal.
The enhanced channeling efficiency of \Ge compared to \Si is experimentally 
demonstrated~\cite{BIINO1997163,doi:10.1063/1.3596709,doi:10.1063/1.4824798}, and it
explains the possibility for \Ge to use shorter crystals with larger bending angles.

The \geant toolkit has also been used to evaluate the deflection efficiency in long crystals with large bending angle 
for negative particles 
in the\tev energy regime. The efficiency is spoiled by the crystal length, resulting in negligibly small efficiencies even at low momentum, as 
shown in Fig.~\ref{fig:geant4_efficiency_dep} for the case of antiprotons traversing a 7\cm long \Si crystal with a 14\mrad bending.
For a  5\cm long \Si crystal with 5\mrad bending 
the efficiency amounts to less than $0.1$\%.

\subsection{Detector occupancy}
\label{sec:occupancy}


The interaction of the impinging protons on the fixed target might represent a challenge for the 
detector operations, radiation hardness, and event reconstruction.

The fluence depends mainly on the average number of primary and secondary interactions taking 
place in the fixed target.
The average number of primary interactions can be determined as $\nu = F p / f$,
where $p$ is the probability for a proton to interact in the target material, $F$ the rate of impinging protons,
and $f$ the \lhc bunch collision frequency, $f=11245~\text{Hz}\times2400=27$~MHz assuming 2400 bunches.
The probability $p$ can be estimated itself as $p = 1-e^{-T/\lambda}$, where $T$ is the thickness of the target material and $\lambda$ its
nuclear interaction length.
For the \W target, $T=0.5\cm$ and $\lambda=9.95\cm$ results in $p=0.05$, which in turn gives $\nu = 0.91$ for a flux $F=5\times10^8$ protons/s.
Similarly, for the \Si (\Ge) crystal with $T=L$ as given in Table~\ref{tab:crysparam} and using $\lambda = 46.52~(26.86)\cm$ 
we obtain $\nu = 2.59~(3.15)$. Summing together the \W and \Si (\Ge) contributions gives $\nu = 3.49~(4.05)$.  
Similar results are obtained evaluating the number of primary interactions as 
$\nu = F N_A \sigma_{\pr\pr} \rho T A_{\rm part} / A_T$,
where $N_A$ is the Avogadro number, $\rho$ the target density, $A_T$ the atomic mass,
$\sigma_{\pr\pr}=48$\mbarn the total \pr\pr cross section at $\sqrt{s}\approx115$\tev,
and $A_{\rm part}$ the number of participant nucleons, estimated using the Glauber model for $\pr$Pb collisions~\cite{Loizides:2016djv} and
rescaling to \W and \Si (\Ge) assuming a spherical geometry for nuclei.
These values are about a factor two smaller than for nominal \pr\pr collisions for the \lhcb upgrade, 
$\nu_{\pr\pr} = {\cal L} \sigma_{\pr\pr}/f = 7.6$, with ${\cal L}=2\times10^{33}$~\invcmtwo\invs and a total \pr\pr cross section
at $\sqrt{s}=14$\tev of $\sigma_{\pr\pr}=102.5$\mbarn.  

\begin{figure}[htb]
\centering
\includegraphics[width=0.4\textwidth]{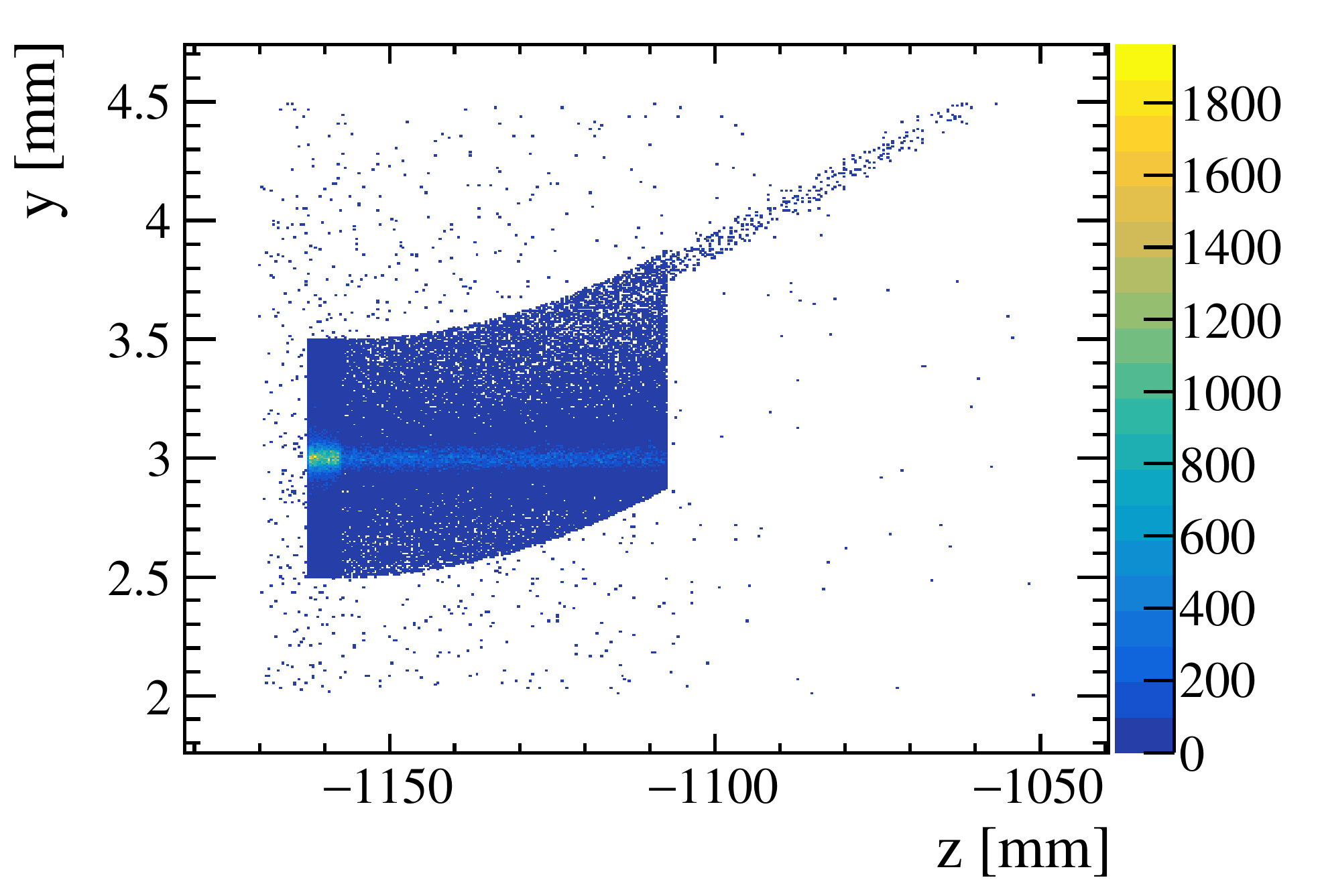}
\caption{\label{fig:radiography} Radiography of the \W target and \Ge crystal geometries in \geant, shown in the $zy$ view.
The distribution represents the origin vertex for stable charged particles from different physics processes 
(Compton, $\delta$ rays, hadronic interactions and pair production). 
The presence of the \Lc decay vertex after the crystal is clearly visible.
}
\end{figure}

Secondary interactions have been studied implementing the geometry of the \W target and \Ge crystal 
of \Sa scenario
 in the \geant framework. 
 Interactions of protons with the \W target and the \Ge crystal are generated
 using \epos~\cite{Pierog:2013ria}, tuned to the corresponding average number of primary interactions, 
whereas \Lc events use~\pythia and \evtgen~\cite{Lange:2001uf} to describe their production and decay.
Figure~\ref{fig:radiography} illustrates the radiography of the device based on the distribution of stable charged particles.
A figure of merit of the occupancy can be obtained from the average number of charged particles within the detector
acceptance, using the simplified geometrical model discussed in Sec.~\ref{sec:layout}. 
This value is found to be similar to the corresponding average in the case of nominal \pr\pr collisions for the \lhcb upgrade.

%
%
\subsection{Characterization of signal events}
\label{sec:characterization}
\begin{figure}[htb]
\centering
\includegraphics[width=0.48\textwidth]{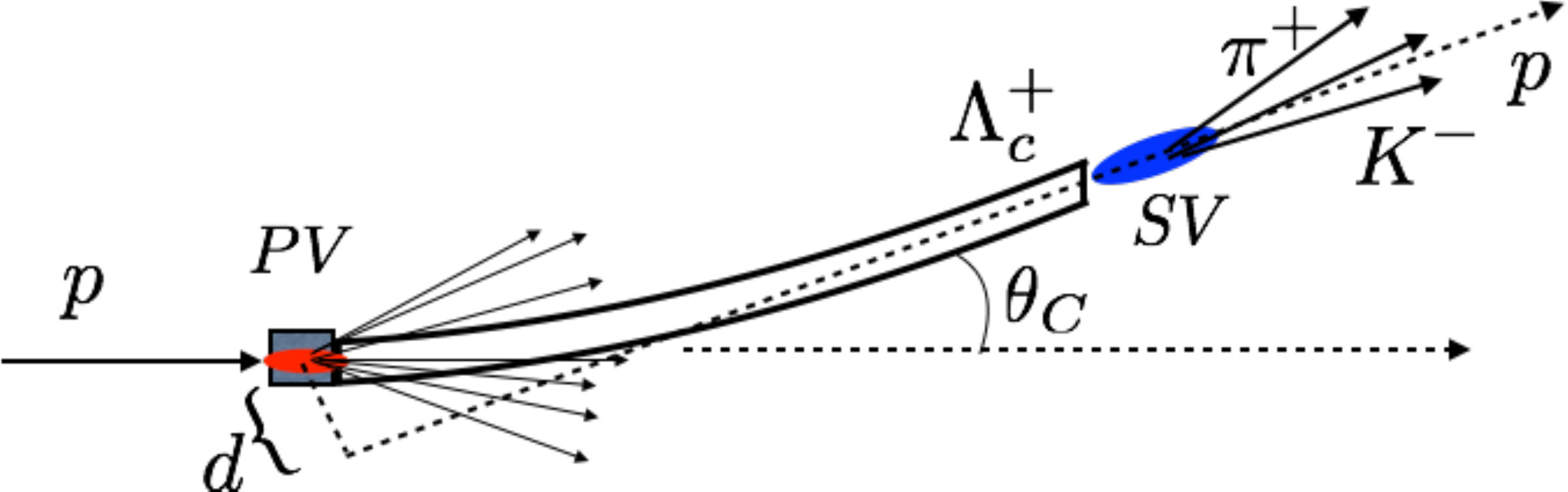}
\caption{\label{fig:SignalSketch} 
Sketch of a signal event: the \Lc baryon, produced in the \W target at the primary vertex (\PV) 
of the event, is channeled and deflected at an angle determined by the crystal curvature $\theta_C$.
The secondary vertex (\SV) of the \Lc baryon decay is outside the crystal, and its position is
determined by the vertex of the decay products.
}
\end{figure}

%
%
A sketch of a \Lc signal event is shown in Fig.~\ref{fig:SignalSketch}.
Particles from fixed-target collisions are emitted in a cone with opening angle $\approx1\mrad$
around  the proton direction.
Only particles entering the crystal with momenta parallel to the atomic planes
within the Lindhard critical angle of few \murad, $\theta_y \in (-\theta_L,\theta_L)$, are channeled and bent inside the detector acceptance.
From the distribution of the polar angle $\theta_y$ versus the momentum, shown in Fig.~\ref{fig:ThetavsP_all},
the fraction of channeled particles is estimated $\approx 10^{-4}$, and represents the hard component
of the momentum spectrum.
Most of the particles produced in the target are not channeled, remaining confined inside the beam pipe.
A small tilt of few  Lindhard angles of the crystal atomic planes with respect to the incoming proton
direction will avoid channeling of non-interacting protons with no cost in trapping efficiency.

Baryon decay products are also highly collimated, and might be difficult to reconstruct by the detector. 
For $\Lc\to\pr\Km\pip$ decays, \pythia simulations indicate that the average angular separation between all pair of
tracks range from 1.9 to 2.7\mrad, depending on their masses. For the \Sa scenario,
this results in an average radial separation at 
the first detection layer of about 1.7\mm, well above the 55\murad pixel pitch~\cite{LHCb-TDR-013}. 
This would limit the target position a few\cm upstream the first sensor in \Sb scenario.

The high momentum, the polar angle, 
and the invariant mass of the outcoming baryons
define a distinct signature of signal events. By applying a few selection criteria based on these kinematic variables, 
\eg the angle $\theta_y$ (invariant mass) to be within a few $\sigma$ of $\theta_C$ (nominal baryon mass) and
the momentum $p \gtrsim 800\gevc$, it would be possible to obtain a high background rejection while 
retaining most of signal events. 
For this purpose, the \lhcb detector performance are crucial:
in particular the estimated track angle resolution of $\approx$25\murad,
combined with good momentum and mass resolutions, $\approx 1\%$~\cite{LHCb-DP-2014-002} and $\approx20$\mevcc~\cite{LHCb-CONF-2017-001}, respectively.
%
Particle identification is very limited at momentum regime of several hundreds \gevc, and is neglected in
 this study. We assign the particle mass hypothesis based on momentum hierarchy, 
\eg for $\Lc\to\pr\Km\pip$, the highest momentum track is assigned to be the \pr, the second the \Km and the third the \pip.

With such criteria the vast majority of the signal candidates are produced in the \W target and 
decay after the crystal, featuring maximal spin precession. 
This is because particles 
produced inside the crystal have lower probability to be channeled, 
and particles decaying inside the crystal are less bent and have smaller polar angles.
The reconstructed invariant mass helps to reject backgrounds from other channeled hadrons
having a similar topology, 
\eg the more abundant $\Dp\to\Km\pip\pip$ and $\Dsp\to\Kp\Km\pip$ decays for $\Lc\to\pr\Km\pip$ signal events.
Information on the primary and decay vertices could also be exploited to further reject baryons
either produced at the beginning or decayed towards the end of the crystal, which might induce a small 
bias on the spin precession.

Before channeling, a net baryon transverse momentum 
is needed to define the 
production plane and have non-zero initial transverse polarization $s_0$, see Fig.~\ref{fig:CrystalPlane}. 
This, in turn, requires a non-zero polar angle $\theta_x$, for which there are no restrictions and has a distribution 
for the hard momentum component similar to that shown in Fig.~\ref{fig:ThetavsP_all}. 
To prevent initial polarization dilution, the baryon spin rotation has to be determined in bins of $\theta_x$.

%
\section{Expected sensitivities}
\label{sec:sensitivity}

\begin{table*}[htb]
\caption{
Production cross sections $\sigma$, initial polarizations $s_0$, nuclear modification factors $R_{\pr T}$, and antibaryon-to-baryon 
ratios \Rq for \pr\W collisions at $\sqrt{s}\approx115\gev$, along with the anomalous magnetic moment $g'=(g-2)/2$, decay channels, 
branching ratios \br and decay asymmetry parameters $\alpha_f$, for the different charm, beauty and strange charged baryons.
For comparison purposes, the \Lc case has been considered in the $\Deltares^{++} \Km$ and $\Lz \pip$ final states. 
Other quantities like particle masses, spins and lifetimes are taken from Ref.~\cite{Olive:2016xmw}.
For \Xibp and \Omegabp antibaryons, \br includes the fragmentation fraction from \bquarkbar quarks, and $\sigma$ is the total 
 $\proton\proton\to\bquark\bquarkbar$ beauty cross section.
\label{tab:quantities1}}
\makebox[\textwidth]
{\begin{tabular}{lccccccc}
\toprule
Particle                     & \multicolumn{2}{c}{\Lc}   & \Xicp     & \Xibp              & \Omegabp           & \Xip        & \Omegap \\
\midrule
Decay channel       & $\Deltares^{++}\Km$ & $\Lz\pip$ & $\Deltares^{++}\Km$ & $\Xip \jpsi$    & $\Omegap \jpsi$    & $\Lbar\pip$ & $\Lbar\Kp$  \\
                    &                    &           &                    & $\Xicbarz \pip$ & $\Omegacbarz \pip$ &             &  \\
Cross section, $\sigma$ [mb] & \multicolumn{2}{c}{$0.0182$}  & $0.0129$  & $4.67\times10^{-3}$ & $4.67\times10^{-3}$ & $3.4$       & $1.03$  \\
$|s_0|$                      & \multicolumn{2}{c}{$\phm0.6$} & $\phm0.6$ & $0.6$              & $0.6$              & $0.5$       & $0.2$   \\
$g'$                         & \multicolumn{2}{c}{$-0.3$}    & $-0.3$    & $1.4$              & $5.8$              & $1.9$       & $2.2$   \\
\br                          & $\phm1.09\%$  & $\phm0.83\%$ & $0.31\%$  & $2.9\times10^{-6}$  & $8.3\times10^{-7}$  & $63.83\%$   & $43.32\%$  \\
$\alpha_f$                   & $-0.67$  & $-0.91$ & $-0.67$   & $0.91$             & $0.91$             & $0.458$     & $-0.642$   \\
$R_{\pr T}$                    & \multicolumn{7}{c}{$\approx 1$} \\
\Rq                          & \multicolumn{2}{c}{$1$}       & $1$     &    $0.5$           &  $0.5$             &     $0.8$     & $0.9$ \\
\bottomrule
\end{tabular}}
\end{table*}

The main contribution to the statistical uncertainty on the $d$ and $g$ factors of \Lc baryons, 
and similarly for all other baryons,
can be estimated in the limit of $\gamma \gg 1$ as
\begin{eqnarray}
&&\sigma_d \approx \frac{g-2}{\alpha_f s_0\left( \cos\Phi -1 \right)}\frac{1}{\sqrt{N_\Lc^{\rm reco}}}~, \nonumber \\
&& \sigma_{g} \approx \frac{2}{\alpha_f s_0 \gamma \theta_C }\frac{1}{\sqrt{N_\Lc^{\rm reco}}}, 
\label{eq:EDM_stat_uncertainty}
\end{eqnarray}
where $N_\Lc^{\rm reco}$ is the number of channeled and reconstructed \Lc baryons. 
These estimates assume negligibly small uncertainties on $\theta_C$, $\gamma$ and $s_0$, and
follow directly from Eqs.~\eqref{eq:EDM_precession} and~\eqref{eq:MDM_precession}.
%
%
%
An alternative approach to assess the sensitivity is to generate and fit pseudo-experiments using a probability density function 
based on the spin precession motion and angular distribution, Eqs.~(\ref{eq:EDM_precession}),~(\ref{eq:MDM_precession}) and~(\ref{eq:Lc_AngDist}).
The two methods provide consistent results, although the former tends to underestimate the uncertainties by about
a factor two compared to the latter.

The number of \Lc baryons, and similarly for all other baryons, channeled in the bent crystal and reconstructed by the 
detector can be estimated as
\begin{equation}
\label{eq:NLcReco}
  N_\Lc^{\rm reco} = N_\Lc \br(\Lc\to f)\effCH\effDF\effdet ,
\end{equation}
where $\br(\Lc\to f)$ is the branching fraction of the \Lc decay into the final state $f$, 
\effCH is the fraction of channeled baryons in the crystal,
\effDF is a ``decay flight'' efficiency that accounts for the fraction of channeled \Lc baryons decaying
after the crystal and within the detector fiducial volume,
and
\effdet is the detector reconstruction efficiency for $\Lc\to f$ decays.
%
%
The number of \Lc baryons produced with 7\tev protons on a fixed target can be estimated as
\begin{equation}
N_\Lc = \frac{F t}{S}\sigma(\proton\proton\rightarrow\Lc X) \Rq N_T ,
\label{eq:lambdac_rate}
\end{equation}
where $F$ is the proton rate, $t$ the data taking time, $S$ the beam transverse area, $N_T$ the number of target nucleons,
$\sigma(\proton\proton\rightarrow\Lc X)$ the cross-section for \Lc production
in \proton\proton interactions at $\sqrt{s}=114.6$\gev center-of-mass energy,
and \Rq is the antibaryon-to-baryon ratio for the case of antibaryon production. 
%
The number of target nucleons is $N_T=N_A\rho S T A_N R_{\pr T} /A_T $,
where $N_A$ is the Avogadro number, $\rho$ ($T$) is the target density (thickness), $A_T$ ($A_N$) is the atomic mass (mass number), 
and $R_{\pr T}$ is a nuclear modification factor taking into account that the number of participant nucleons of the target nuclei, $A_{\rm part}$,
differs from $A_N$ due to nuclear matter effects.
For hard processes in the absence of strong final-state interaction $A_{\rm part}$ scales with $A_N$~\cite{Loizides:2016djv}, 
thus we take $R_{\pr T} \approx 1$.
For the tungsten target $\rho=19.25$~g/cm$^3$, $A_T=183.84$ g/mol, $A_N=183.84$, and $T=0.5$\cm.

All the necessary inputs and their values as used for the sensitivity study,
summarised in Tables~\ref{tab:quantities1} and~\ref{tab:quantities2},
are taken from a combination of measurements, estimates and Monte Carlo simulations, 
and are discussed in detail in the following, along with the final results. 

\subsection{Baryon and antibaryon production yields}

The \Lc and \Xicp baryon cross sections can be estimated from the total charm production cross section measured by the 
PHENIX experiment in $\pr\pr$ collisions at $\sqrt{s} = 200\gev$~\cite{Adare:2006hc}
rescaled to $\sqrt{s} = 114.6 \gev$, assuming a linear dependence on $\sqrt{s}$,
and the corresponding fragmentation fractions. 
For the \Lc case the fragmentation fraction $f_\Lc \approx 5.6\%$ is derived from~\cite{Adare:2006hc},
consistent with theoretical predictions~\cite{Kniehl:2005de}. 
%
The \Lc baryon branching fractions to $\Deltares^{++}\Km$ and $\Lz\pip$ final states are taken from Ref.~\cite{Olive:2016xmw}.
The \Xicp fragmentation fraction is estimated considering that all known \cquark-hadron fractions, which amount to about 92\%,
leave room for the unknown \Xicp, \Xicz and \Omegacz fractions~\cite{Lisovyi:2015uqa,Gladilin:2014tba}. 
Assuming $f_\Xicp \approx f_\Xicz \gg f_\Omegacz$, we obtain $f_\Xicp \approx 4\%$, which is used to rescale 
by a factor $f_\Xicp/f_\Lc \approx 0.71$ the \Lc cross section.
The absolute $\Xicp\to\Deltares^{++}\Km$ branching fraction is estimated from $\br(\Xicp\to\proton\Km\pip)$, 
measured relative to that of $\Xicp\to\Xim\pip\pip$, considering that all known decay modes sum to the total width and assuming that 
the relative resonant contribution to the $\Xim\pip\pip$ final state is the same in \Xicp and \Lc decays.
%
No other quasi-two body \Lc or \Xicp decays to the final state $\pr \Km \pip$ are considered for this study.
However, there are additional contributions, \eg $\Lc\to\Kstarzb\pr$ and $\Lc\to\Lambda(1520)\pip$, with 
similar branching fractions~\cite{Olive:2016xmw,Aitala:1999uq} that can be exploited to improve the
sensitivity.


For \Xibp and \Omegabp baryons produced from 7\tev protons impinging on fixed target, the total beauty cross section can be estimated by
rescaling the $\pr\pr\to\bbbar$ cross section measured at $\sqrt{s} = 7\tev$~\cite{Aaij:2010gn}.
%
%
%
As a working hypothesis the ratios \Rq for bottom baryons are assumed to be $\approx 0.5$,
on the basis of the results for charm hadron production
at lower energies~\cite{Braaten:2002yt,Anjos:2001jr,Herrera:1997qh}.
Branching fractions for \Xibp baryons are known for very few final states. Two suitable two-body decays, requiring a simple two-body angular 
analysis, are considered. Firstly, the $\Xibp\to \Xip \jpsi$ decay, where the \jpsi and \Xip can be detected in the dimuon final state and as
a positive track, respectively. This decay has been measured and its branching fraction times the \Xibp fragmentation function is $\approx 6 \times 10^{-7}$~\cite{Olive:2016xmw}.
%
%
Secondly, the $\Xibp \to \Xicbarz \pip$ decay, where the charm antibaryon can be reconstructed
in the $\Xip \pim\pip\pim$,  $\Xip \pim$ or $\antipr \Kp \Kp \pim$ final states.
%
%
This decay has not been observed but its branching fraction can be estimated by comparing 
the efficiency corrected signal yields for $\Xibm \to \Xicz \pim$ and $\Lb \to \Lc \pim$ decays~\cite{Aaij:2014lxa}.
The average fragmentation fraction $f_\Lb \approx 7\%$~\cite{Amhis:2016xyh},
the measured $\br(\Lb \to \Lc \pim)$~\cite{Olive:2016xmw}, and the \Xicz branching ratios,
estimated similarly to the previous \Xicp case, are used for this calculation.
%
%
Summing together the contributions of the $\Xibp\to \Xip \jpsi$ and the $\Xibp \to \Xicbarz \pip$ decays,
we obtain a global branching fraction times the fragmentation function as 
shown in Table~\ref{tab:quantities1}.
Similar decays can be considered for the \Omegabp baryon. 
In this case, the $\Omegabp \to \Omegacbarz \pip$ and $\Omegacbarz$ branching ratios are unknown,
and we assume  the latter to be the same as for \Xicz decays and scale $f_\Omegabp \br(\Omegabp \to \Omegacbarz \pip)$ by $\approx 0.29$, from the ratio between $f_\Omegabm \br(\Omegabm \to \Omegam \jpsi)$ and $f_\Xibm \br(\Xibm \to \Xim \jpsi)$~\cite{Olive:2016xmw}.
%

\begin{table*}[htb]
\caption{Channeling, survival, decay flight and detector efficiencies, along with the average energy and
squared transverse momentum of channeled baryons (before decay flight requirements), 
for a \W target with a \Si or \Ge bent crystal in the $\Sa$ $[\Sb]$ scenario.
Note that \effDF already includes \effs.
The sensitivity study based on pseudo-experiments makes use of the complete energy spectrum after channeling, from which
$\overline E$ and $\overline{p_\perp^2}$ reported here are obtained.
All estimates, except \effdet (see text), are obtained from samples of charm and strange baryons generated separately
for each baryon type from 7\tev proton beam collisions on protons at rest using \pythia. 
For beauty baryons, we scale the energy of other simulated baryons to obtain an average energy shift 
estimated assuming a linear dependence with the baryon mass difference.
\label{tab:quantities2}}
\makebox[\textwidth]
{\begin{tabular}{lccccccc}
\toprule
Particle      & \multicolumn{2}{c}{\Lc}  & \Xicp    & \Xibp   & \Omegabp & \Xip   & \Omegap     \\
\midrule
Decay channel  & $\Deltares^{++}\Km$  & $\Lz\pip$ & $\Deltares^{++}\Km$ & $\Xip \jpsi$    & $\Omegap \jpsi$    & $\Lbar\pip$ & $\Lbar\Kp$  \\
               &                     &           &                    & $\Xicbarz \pip$ & $\Omegacbarz \pip$ &             &             \\
\cmidrule{2-8}
               & \multicolumn{7}{c}{Si} \\
\effCH $[\times10^{-4}]$    & \multicolumn{2}{c}{$1.24~[4.14]$} & $1.04~[3.90]$  & $2.09~[8.91]$  & $2.11~[9.10]$  & $1.75~[5.57]$  & $1.44~[3.84]$ \\
$\overline E$ $[\tev]$     & \multicolumn{2}{c}{$1.36~[2.70]$} & $1.24~[2.40]$  & $1.24~[2.44]$  & $1.24~[2.48]$  & $1.12~[1.54]$  & $1.09~[1.33]$  \\
$\overline{p_\perp^2}$ $[\gevgevcc]$ & \multicolumn{2}{c}{$1.22~[0.75]$} & $1.09~[1.19]$  & $1.55~[1.25]$  & $1.49~[1.25]$ & $0.20~[0.21]$ & $0.34~[0.32]$ \\
%
\effs  $[\%]$              & \multicolumn{2}{c}{$9.9~[6.9]$} & $31.7~[24.9]$  & $46.3~[41.0]$  & $45.1~[40.1]$  & $99.8~[99.8]$  & $99.5~[99.3]$ \\ 
\effDF $[\%]$              & $9.9~[6.9]$       & $0.42~[0.16]$     & $31.7~[24.7]$  & $46.3~[39.5]$  & $45.0~[38.7]$  & $0.08~[0.05]$  & $0.20~[0.15]$ \\
\cmidrule{2-8}
              & \multicolumn{7}{c}{Ge} \\
\effCH $[\times10^{-4}]$     & \multicolumn{2}{c}{$2.32~[5.57]$} & $2.06~[5.18]$  & $3.92~[11.34]$  & $3.98~[11.63]$  & $3.18~[7.34]$  & $2.57~[5.17]$ \\
$\overline E$ $[\tev]$      & \multicolumn{2}{c}{$1.37~[2.26]$} & $1.30~[2.07]$  & $1.31~[2.16]$  & $1.32~[2.18]$  & $1.19~[1.51]$  & $1.14~[1.30]$  \\
$\overline{p_\perp^2}$ $[\gevgevcc]$ & \multicolumn{2}{c}{$1.16~[1.05]$} & $1.47~[1.09]$ & $1.51~[1.32]$ & $1.52~[1.33]$  & $0.22~[0.22]$  & $0.35~[0.33]$ \\
%
\effs $[\%]$                & \multicolumn{2}{c}{$20.0~[17.4]$} & $44.9~[40.9]$  & $59.0~[57.4]$  & $57.9~[56.5]$  & $99.9~[99.9]$  & $99.7~[99.6]$ \\ 
\effDF $[\%]$               & $20.0~[17.4]$       & $0.85~[0.52]$     & $44.9~[40.7]$  & $58.9~[56.0]$  & $57.8~[55.2]$  & $0.08~[0.06]$  & $0.20~[0.16]$ \\
\cmidrule{2-8}
\effdet $[\%]$ & $20$               & $10$      & $20$               & $12$            & $12$               & $10$        & $10$        \\
\bottomrule
\end{tabular}}
\end{table*}

For the \Xip and \Omegap antibaryons, which contain two and three \squark valence antiquarks respectively,
the cross sections are estimated by scaling the \Lz production cross section using the universal 
strangeness suppression factor at high energies, $\lambda_s \approx 0.32$~\cite{Arakelyan:2015nua}.
In turn, the \Lz production cross section is estimated from the inclusive $\pr\pr\to \Lz X$ cross section measured 
at beam momenta of 158\gev~\cite{Aduszkiewicz:2015dmr} and 405\gev~\cite{Kichimi:1979te} 
($\sqrt{s}\approx 17.2$ and $27.6$\gev, respectively).
%
The ratios \Rq are taken to be $0.8$ and $0.9$ for \Xip/\Xim, \Omegap/\Omegam, respectively,
as inferred from Au+Au collisions at $\sqrt{s_{NN}}=130$\gev~\cite{Adams:2002pf}.
All branching ratios are in this case known~\cite{Olive:2016xmw}.

\subsection{Efficiencies}

The channeling efficiency \effCH in \Si and \Ge crystals includes both the trapping efficiency \efft and deflection efficiency \effc, 
and has been estimated separately for each baryon type following the procedure 
described in Sec.~\ref{sec:crystalparams}. 
The trapping efficiency itself accounts for the angular and momentum divergence of the baryons produced in the target, 
and is evaluated from the fraction of baryons within the Lindhard angle and momentun $\gtrsim 800\gevc$.
%
Crystal parameters, optimized for charm baryons, are taken to be common for all baryon species.


%
%
The decay flight efficiency \effDF has two contributions: the survival efficiency, \effs, which accounts for the fraction of channeled
baryons decaying after the crystal, and the probability \effl for long-lived baryons to decay within the VELO region,
$\approx 80$\cm downstream of the nominal \pr\pr collision point.
%
%
When one of the baryon decay products is a long-lived \Lz, \effl also accommodates the probability of the \Lz to decay 
before the large-area tracking system upstream the magnet, $\approx 2$\m downstream of the collision point,
assuming it takes on average half of the initial baryon momentum. 
For simplicity, the same requirements are applied for both \Sa and \Sb scenarios.

The detector efficiency \effdet can be estimated from the product of the geometrical, trigger and tracking efficiencies,
the latter including combinatorics and selection efficiencies. 
The software-based trigger for the \lhcb upgrade detector~\cite{LHCb-TDR-016}, our \Sa scenario, is expected to have efficiency for 
charm hadrons 
comparable to the current high level trigger~\cite{LHCb-DP-2014-002},
$\approx 80\%$, and similarly for other baryons. 
A specific trigger scheme for the fixed-target experiment based on the distinct signature of the signal
events can enhance the trigger efficiency 
to $\approx 100\%$.
%
The tracking efficiency is estimated to be 70\% per track.  
%
Following the discussion in Sec.~\ref{sec:setup}, the geometrical efficiency is taken $\approx 50\%$.
%
For decays to final states including \Lz baryons we further apply a penalty factor $1/2$ to account for the additional inefficiencies to 
reconstruct highly displaced vertices. Note that the inefficiency due to the long lifetime of the
\Lz baryon is separately taken into account in \effDF, as discussed before.
%

\subsection{Spin polarization of baryons}


The asymmetry parameter of the $\Lc\rightarrow \Lz\pip$ decay has been measured to be
$-0.91\pm0.15$~\cite{Olive:2016xmw}.
For other \Lc decays no measurements are available but an effective $\alpha_f$ parameter
can be calculated from a Dalitz plot analysis of $\Lc\to\proton\Km\pip$
decays~\cite{Aitala:1999uq}, \eg $\alpha_{\Lc\to\Deltares^{++}\Km} = -0.67\pm0.30$~\cite{Botella:2016ksl}.
Eventually, a Dalitz plot analysis of $\Lc\to\proton\Km\pip$ decays
would provide the ultimate sensitivity to dipole moments.
For the $\Xicp\to\Deltares^{++}\Km$ decay the asymmetry parameter is taken to be similar to the \Lc decay to the same final state,
whereas for all beauty antibaryon decays it is assumed to be about the same as for the $\Lc\to\Lz\pip$ decay.
For the $\Xip\to\Lbar\pip$ decay the asymmetry parameter is taken from~\cite{Olive:2016xmw}.
The $\Omegap\to\Lbar\pip$ decay is predominantly parity-conserving and thus has a negligibly small asymmetry 
parameter~\cite{Chen:2005aza}. The polarization can be determined in this case from the angular distribution of the antiproton from 
the \Lbar decay, as the \Omegap polarization can be related to the polarization of its \Lbar child baryon such 
that ${\mathbf s}_{\Omegap} = {\mathbf s}_{\Lbar}$~\cite{Woods:1996hq,Luk:1992ku}.

The initial polarization of \Lc particles produced from the interaction of 7\tev protons on a fixed target is unknown. 
However, a measurement from interaction of 230 \gevc~\pim on copper target 
yields $s_0=-0.65^{+0.22}_{-0.18}$ for \Lc transverse momentum larger than 1.1\gevc~\cite{Jezabek:1992ke}.
Moreover, from data produced in the interactions of 500~\gevc $\pim$ on five thin target foils (one platinum,
four diamond)~\cite{Aitala:1999uq},
the polarization of the \Lc is measured as a function of the \Lc transverse momentum. 
The average polarization is about $-0.1$, reaching $-0.7$ for $p_\perp^2$ between about 1.24 and 5.2$\gev^2/c^2$.
Considering these measurements and the average transverse momentum of channeled \Lc baryons given in 
Table~\ref{tab:quantities2}, 
we assume $|s_0|=0.6$ for both \Lc and \Xicp baryons~\cite{Samsonov:1996ah}.
The same polarization is assumed for the \Xibp and \Omegabp antibaryons.
%
%
Similarly, the initial polarization of \Xip and \Omegap antibaryons produced from the interaction of 7\tev protons on fixed target
is unknown. From proton production below the\tev region~\cite{Woods:1996hq,Luk:1992ku,Ho:1990dd,Trost:1989mq,Rameika:1986rb,Heller:1983ia},
the \Xip are found to be polarized with the same sign and magnitude as the \Xim, increasing about linearly with momentum 
and reaching $\approx -0.2$ at 250\gevc, whereas the \Omegam is consistent with no polarization.
%
As a working hypothesis initial polarizations of $|s_0|=0.5$ and $0.2$ are assumed for \Xip and \Omegap, respectively, considering the large momentum of channeled antibaryons $\approx 1\tev$/c.

Theoretical predictions of $g-2$ for the \Lc and \Xicp baryons range between
$-0.64$ and $0.22$~\cite{Sharma:2010vv,Samsonov:1996ah}, thus a central value $g'=(g-2)/2 =-0.3$ is considered.
For \Xibp and \Omegabp antibaryons we take effective quark mass MDM calculations~\cite{Dhir:2013nka}.
For all strange baryons under consideration there exist measurements~\cite{Olive:2016xmw}.

\subsection{Results}

\begin{figure}[htb]
\centering
\includegraphics[scale=0.35]{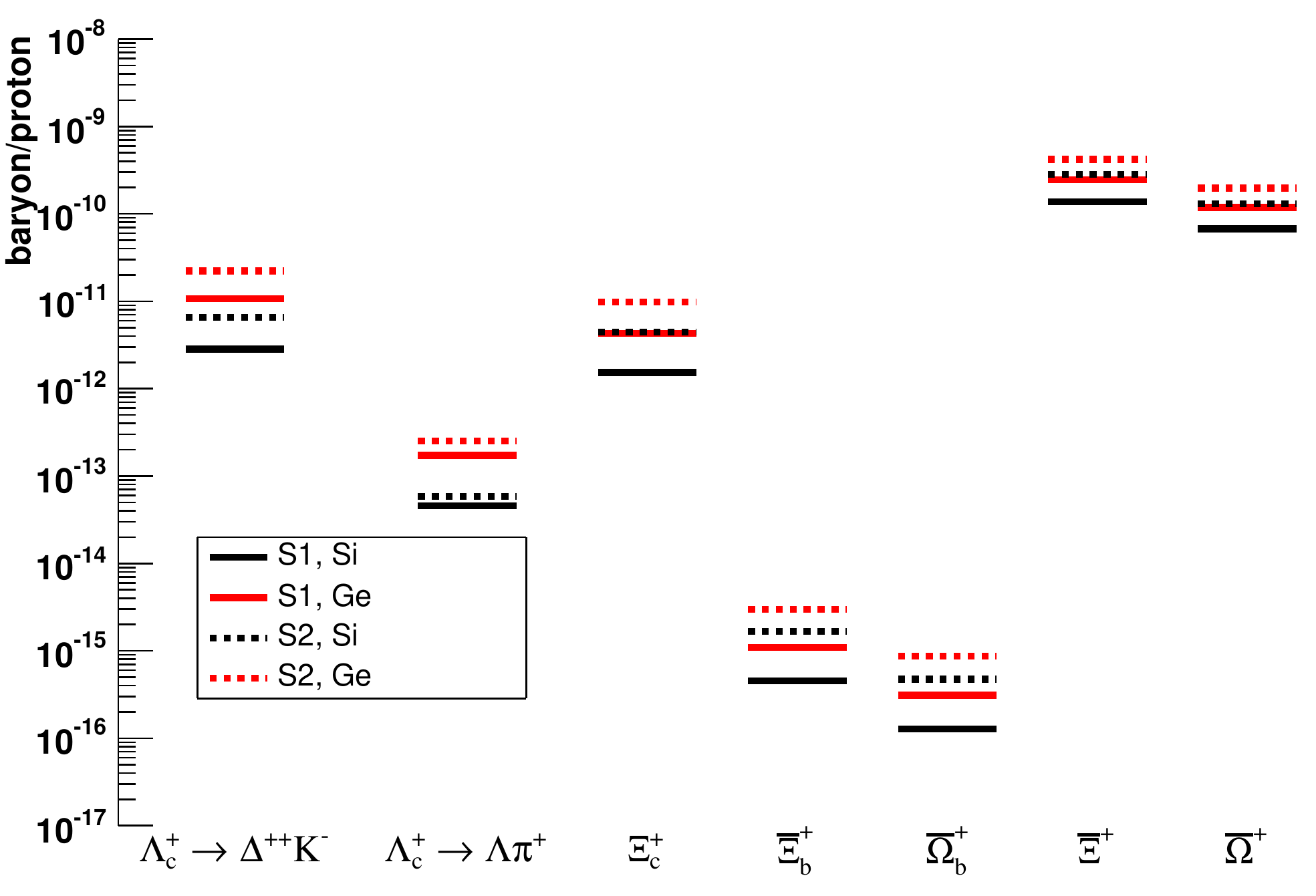}
\vskip0.5cm
\includegraphics[scale=0.35]{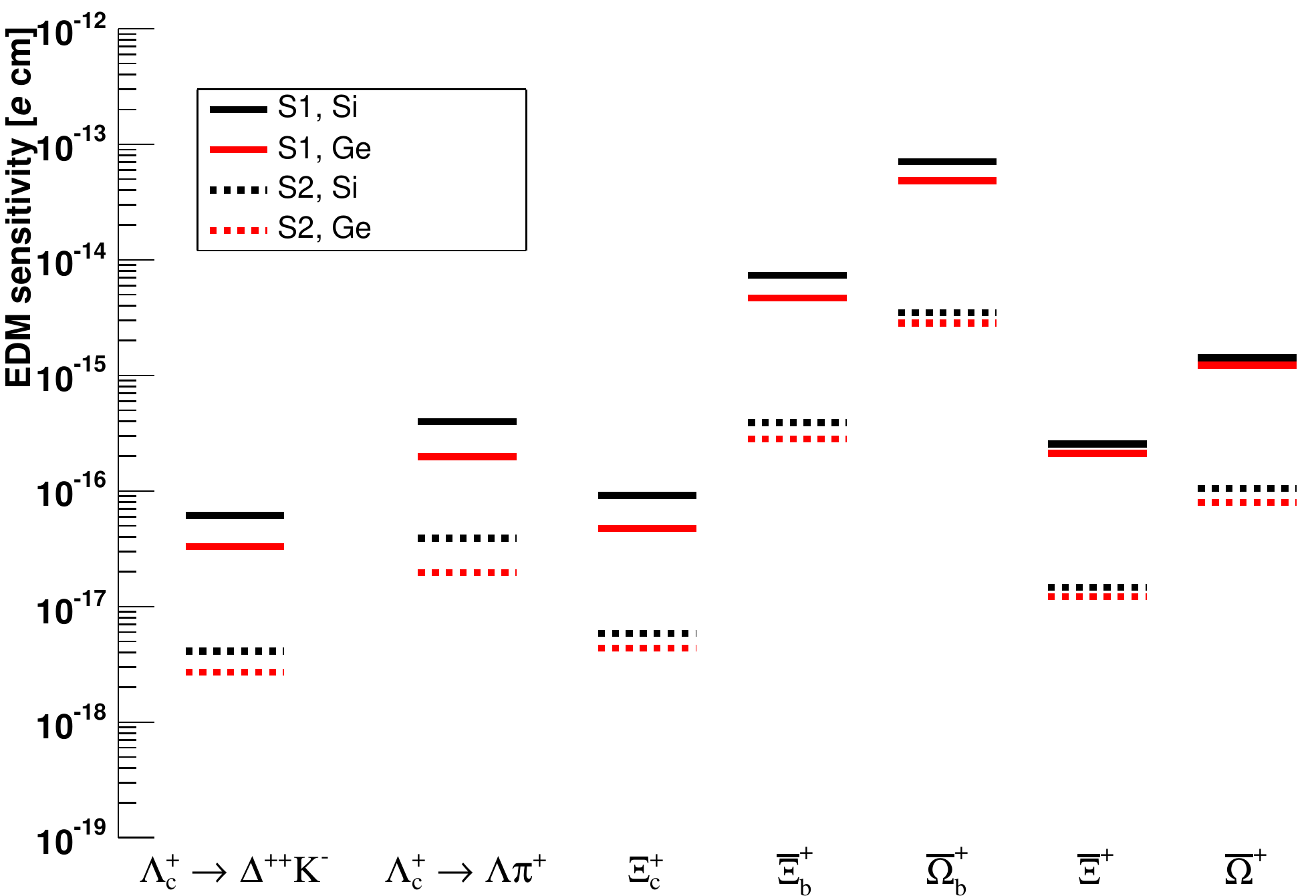}
\vskip0.5cm
\includegraphics[scale=0.35]{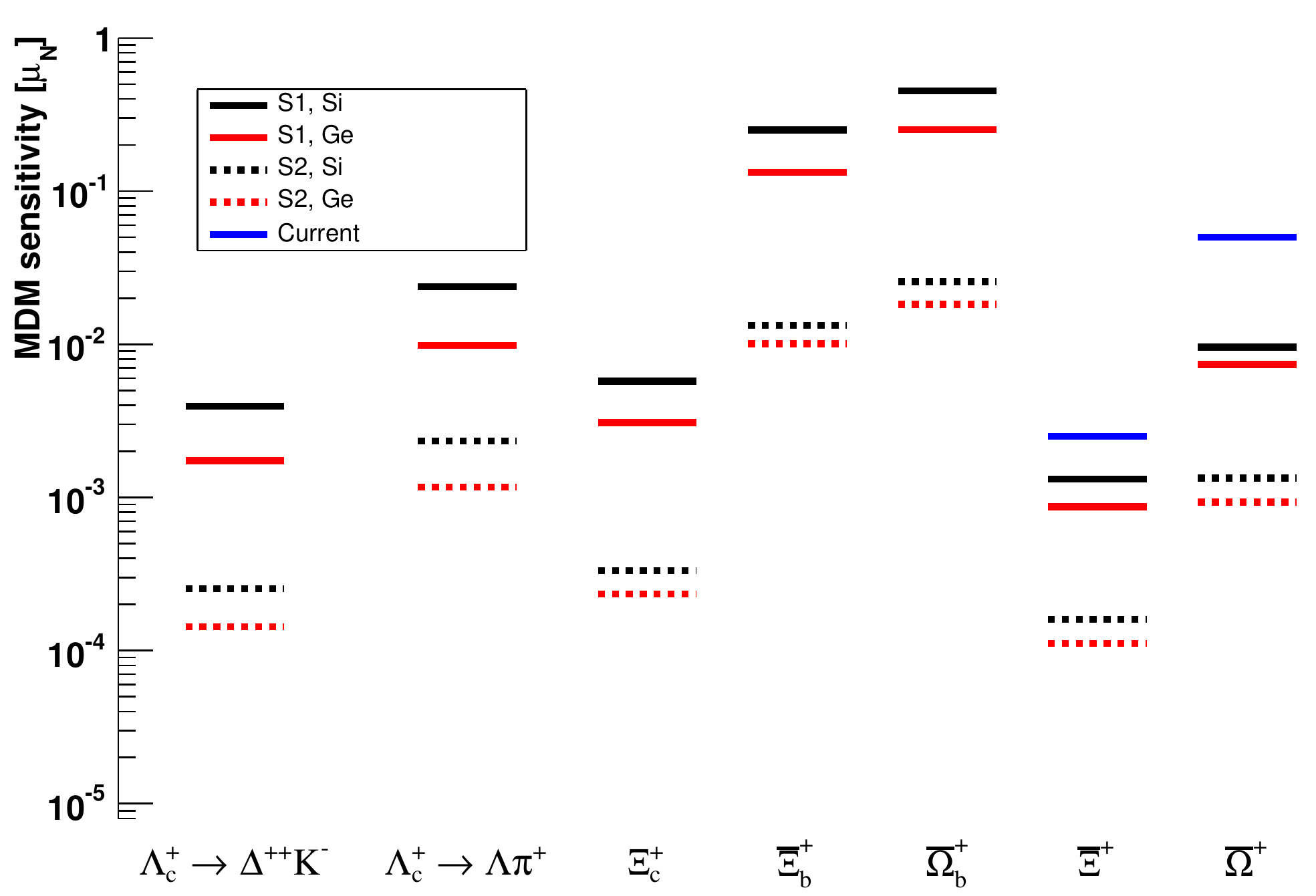}
\caption{
(Top) Estimated yields of channeled and reconstructed signal baryons per incident proton on target,
and (middle) EDM and (bottom) MDM sensitivities,
for \Si and \Ge with crystal parameters optimized for \Sa and \Sb scenarios.
A total of $10^{15}$ and $10^{17}$ protons on target have been considered for \Sa and \Sb respectively.
For comparison purposes, the \Lc case has been studied in the $\Deltares^{++} \Km$ and $\Lz \pip$ final states. 
Blue lines show the sensitivity of the current \Xim and \Omegam MDM measurements.
\label{fig:phys_reach_nbaryons_mu_delta}}
\end{figure}

Combining all parameters, measurements and estimates discussed above and summarized in 
Tables~\ref{tab:quantities1} and~\ref{tab:quantities2},
we obtain the signal yields, 
normalized to the incident proton flux $F$, shown in (top) Fig.~\ref{fig:phys_reach_nbaryons_mu_delta}.
%
These rates procure the expected EDM and MDM sensitivities reported in (middle and bottom) Fig.~\ref{fig:phys_reach_nbaryons_mu_delta},
for both \Si and \Ge crystals and the two considered experimental scenarios, \Sa with $10^{15}\pot$ and \Sb with $10^{17}\pot$.
Germanium crystals provide in all cases significantly better EDM (MDM) sensitivities,
which are for \Sa scenario of order 
 $10^{-17}$, $10^{-14}$ and $10^{-16}$~$e\cm$ 
($10^{-3}$, $10^{-1}$ and $10^{-3}$~$\mu_N$)
for charm, beauty and strange baryons, respectively. 
Here $\mu_N=e\hslash/2m_pc$ is the nuclear magneton, and $m_p$ the proton mass.
Sensivities for \Sb scenario would improve by about one order of magnitude.

%
\section{Conclusions}
\label{sec:conclusion}

Electric and magnetic dipole moments of short-lived baryons are powerful probes for physics within and beyond
the SM. However, EDM and MDM for charm and beauty baryons have not been accessible to date.
A unique opportunity to measure at \lhc the EDM and MDM of charm, beauty and strange charged baryons 
has been discussed here. The experimental setup is based on a fixed-target to be installed in the \lhc where protons from the beam halo are 
deflected using a bent crystal, producing transversally polarized baryons from their interactions with the target.
A second bent crystal is positioned after the target where charged baryons that are channeled deflect their trajectory and enter the 
detector acceptance while rotating their spin. The MDM and EDM information can be inferred from 
the measurement of the spin-polarization vector after the crystal by analysing the angular distribution of the baryon decay products.

The planar channeling efficiency for multi-\tev particles and the spin precession in bent crystals is studied using \geant simulations.
The main result is that both positive and negative particles feature spin precession
and the results agree with predictions based on analytical calculations,
also discussed in Ref.~\cite{Botella:2016ksl}.
However, planar channeling efficiency for negative particles is consistently lower than for positive particles,
thus much higher statistics is required to perform useful measurements.
In that case \CPT tests based on the MDM for baryons and antibaryons could be performed.
The possibility of exploiting axial channeling of negative particles has been briefly discussed but more studies are needed, 
including Monte Carlo \geant simulations, before drawing any conclusion on the possibility to measure
electromagnetic dipole moments. 

A program of EDM and MDM measurements for \Lc, \Xicp charm baryons,
\Xibp, \Omegabp beauty antibaryons, and \Xip, \Omegap strange antibaryons, has been discussed.
The feasibility of the experiment based on the \lhcb detector has been assessed relying on
both parameterised and \geant simulations along with a geometrical model of the detector.
Sensitivities for $10^{15}$\pot could be reached within a few weeks of dedicated detector operations
spanned over several years at a flux of $5\times 10^8\proton/\sec$.
The possibility of a dedicated experiment a covering larger pseudorapidity region,
able to afford higher proton fluxes and longer data taking periods, has also been discussed.
For the \lhcb layout, optimal bent crystal parameters are determined to be 7\cm (5\cm) length and 14\mrad (15\mrad)
bending angle 
for \Si (\Ge), whereas for the dedicated experiment are found to be 12\cm (7\cm) and 7\mrad (8\mrad).
In all cases, \Ge crystals provide enhanced sensitivity.

This unique physics program would provide important experimental anchor points
for QCD calculations and searches for physics beyond the SM.
In the case of charm and beauty baryon EDM the limits would be better than current indirect bounds based on the neutron EDM~\cite{Sala:2013osa,Grozin:2009jq,CorderoCid:2007uc},
extending the new physics discovery potential of the \lhc.



%
\section*{Acknowledgements}
 
\noindent 
We express our gratitude to our colleagues of the \lhcb collaboration, in particular to 
M.~Ferro-Luzzi, 
G.~Graziani, 
R.~Lindner,
G.~Passaleva,
M.~Pepe-Altarelli, 
P.~Robbe,
V.~Vagnoni,
G.~Wilkinson, 
for stimulating discussions and very useful feedback.
We acknowledge support from INFN (Italy), MINECO and GVA (Spain), and the ERC Consolidator Grant CRYSBEAM G.A.615089.

\appendix
\section{Discrete ambiguities}
\label{app:ambiguities}

From Eqs.~(\ref{eq:Phi}),~(\ref{eq:EDM_precession}) and~(\ref{eq:MDM_precession}) we observe that, 
if all the three components of the final polarization vector $\mathbf{s}$ are measured, the $g$ and $d$ factors 
can be extracted along with the initial polarization $s_0$, up to discrete ambiguities.
If $\{s_0,~g',~d\}$ is a solution, where $g'=(g-2)/2$ is the anomalous magnetic moment, then 
\begin{equation}
\left\{ -s_0, ~g'\pm \frac{n\pi}{\gamma\theta_C}, 
        ~d\left(1\pm \frac{n\pi}{\gamma\theta_C}\frac{1}{g'}\frac{\cos\Phi-1}{\cos\Phi+1}\right) \right\},~\nonumber\\
\end{equation}
\begin{equation}
\left\{  s_0, ~g'\pm \frac{m\pi}{\gamma\theta_C}, ~d\left(1\pm\frac{m\pi}{\gamma\theta_C}\frac{1}{g'}\right)  \right\}~,
\end{equation}
are also solutions, with $n$ ($m$) an odd (even) integer.
Performing a simultaneous fit in bins of $\gamma$ to the angular distribution described in Eq.~(\ref{eq:Lc_AngDist}), will resolve the ambiguity. 

%
%
 \bibliographystyle{spphys}
 \bibliography{main}
\end{document}